\documentclass[journal]{IEEEtran}

\usepackage{cite}
\usepackage{graphicx}
\usepackage[cmex10]{amsmath}
\usepackage[justification=centering]{caption}
\usepackage{float}
\usepackage{verbatim}
\usepackage{relsize}
\usepackage{amsfonts}
\usepackage{amsthm}
\usepackage{amssymb}
\usepackage{latexsym}
\usepackage{url}
\usepackage{epstopdf}
\usepackage{enumerate}
\usepackage{mathrsfs}
\usepackage{tabulary}
\usepackage{color}
\usepackage{balance}
\usepackage{amsmath}
\usepackage{cases}
\usepackage{caption}

\DeclareMathOperator*{\argmin}{arg\,min}
\newtheorem{theorem}{Theorem}

\newtheorem{corollary}{Corollary}
\newtheorem{proposition}{Proposition}

\newcommand{\bfc}{{\boldsymbol c}}

\newcommand{\one}{{\boldsymbol 1}}

\newcommand{\TRC}{{\mathrm{TRC}}}
\newcommand{\GV}{{\mathrm{GV}}}
\newcommand{\UB}{{\mathrm{UB}}}

\newcommand{\RCE}{{\mathrm{RCE}}}

\newcommand{\LP}{{\mathrm{LP}}}

\newcommand{\sC}{{\mathscr C}}

\newcommand{\dotleq}{~\dot{\le}~}
\newcommand{\dotgeq}{~\dot{\ge}~}

\begin{document}
		
\title{Error Exponent Bounds for the\\ Bee-Identification Problem}

\author{Anshoo Tandon,~\IEEEmembership{Member,~IEEE}, Vincent Y.\ F.\ Tan,~\IEEEmembership{Senior Member,~IEEE}, \\ and Lav R.\ Varshney,~\IEEEmembership{Senior Member,~IEEE}% <-this % stops a space
	\thanks{A.~Tandon is with the Department of Electrical and Computer Engineering, National University of Singapore, Singapore 117583 (email: anshoo.tandon@gmail.com.}%
	\thanks{V.~Y.~F.~Tan is with the Department of Electrical and Computer Engineering, and with the Department of Mathematics, National University of Singapore, Singapore (email: vtan@nus.edu.sg).}%
	\thanks{L.~R.~Varshney is with the Coordinated Science Laboratory and the Department of Electrical and Computer Engineering, University of Illinois at Urbana-Champaign, Urbana, IL~61801 USA (email: varshney@illinois.edu)}}

\maketitle

\begin{abstract}
Consider the problem of identifying a massive number of bees, uniquely labeled with barcodes, using noisy measurements. We formally introduce this ``bee-identification problem'', define its error exponent, and derive efficiently computable upper and lower bounds for this exponent. We show that joint decoding of barcodes provides a significantly better exponent compared to separate decoding followed by permutation inference. For low rates, we prove that the lower bound on the bee-identification exponent obtained using typical random codes (TRC) is strictly better than the corresponding bound obtained using a random code ensemble (RCE). Further, as the rate approaches zero, we prove that the upper bound on the bee-identification exponent meets the lower bound obtained using TRC with joint barcode decoding.
\end{abstract}

%\section{Introduction: The Bee-Identification Problem}
\section{Introduction}
\begin{figure}[h]
	\centering
	\includegraphics[width=0.49\textwidth]{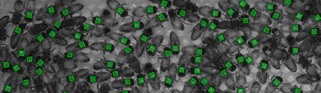}
	\caption{Bees tagged with barcodes (adapted from~\cite{Gernat18}).}
	\label{Fig:Beehive}
\end{figure}

Consider a group of $m$ different bees, in which each bee is tagged with a unique barcode for identification purposes in order to understand interaction patterns in honeybee social networks~\cite{Gernat18}. Assume that a camera is employed to picture the beehive to study the interactions among bees. The image output (see Fig.~\ref{Fig:Beehive}) can be considered as a noisy and unordered set of $m$ barcodes. We formally pose the problem of bee-identification from a beehive image as an information-theoretic problem (Sec.~\ref{Sec:ProbFormulation}).

The bee-identification problem has applications in identification of warehouse products (labeled with unique RFID barcodes) using wide-area sensors. Other applications include package-distribution to recipients from a batch of deliveries with noisy address labels, and similar ``bipartite matching'' settings. It also has potential applications in identification of the mapping between signals and their meaning in ``alien communication'' with extraterrestrials, and also in learning communication protocols among robots, via the use of pilot signals going through the alphabet. 

We consider the scenario where the barcode for each bee is represented as a binary vector of length $n$, and the bee barcodes are collected in a codebook $C$ comprising $m$ rows and $n$ columns, with each row corresponding to a bee barcode. As shown in Fig.~\ref{Fig:BeeChannel}, the channel first permutes the rows of $C$ with a random permutation $\pi$ to produce $C_\pi$. The entries of $C_\pi$ are then subjected to noise (corresponding to a binary symmetric channel (BSC) with crossover probability $p$), and the channel output is denoted $\tilde{C}_{\pi}$.  We assume that the decoder has knowledge of codebook $C$, and its task is to \emph{recover the row-permutation} $\pi$ introduced by the channel. Note that the permutation $\pi$ directly ascertains the identity of all the bees.

\subsection{Related Work}
In a related work motivated by an Internet of Things (IoT) setting, the identification of users in strongly asynchronous massive access channels was studied~\cite{Shahi18MAC}. The  identification of the underlying distributions of a set of observed sequences (where each sequence is generated i.i.d. by a distinct distribution) was analyzed in~\cite{Shahi18}.
The bee-identification problem, on the other hand, allows codebooks where all barcode sequences are generated using the same underlying distribution. 

\begin{figure}[t]
	\centering
	\includegraphics[width=0.45\textwidth]{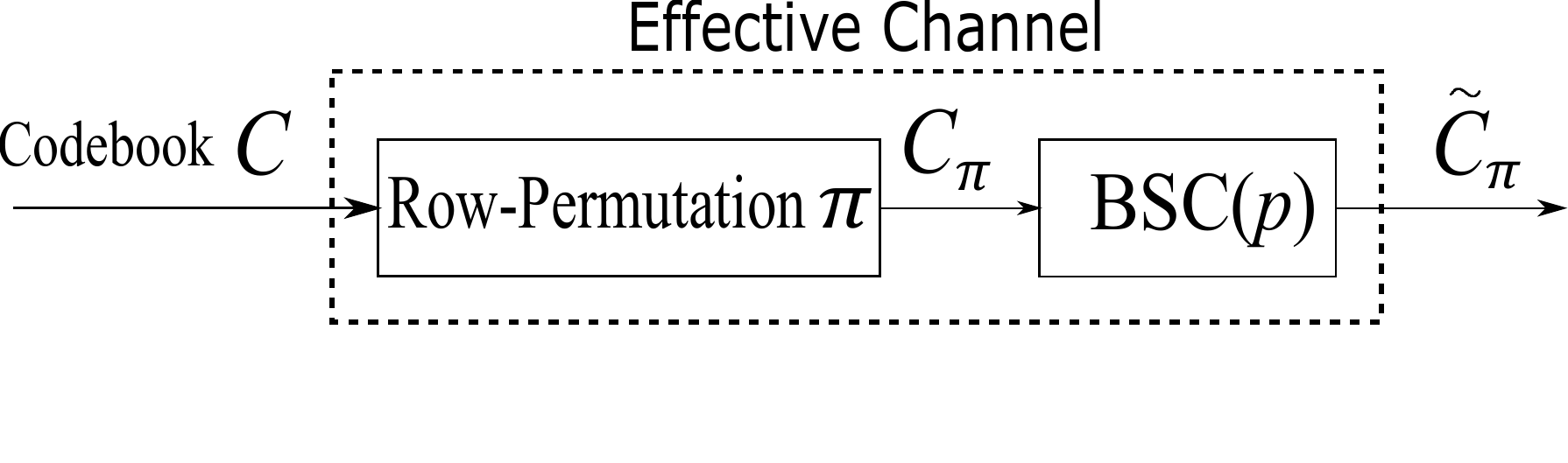}
	\caption{Effective channel for the bee-identification problem.}
	\label{Fig:BeeChannel}
\end{figure}

In another related work \cite{Heckel17}, the fundamental limits of data storage via unordered DNA molecules was investigated. Here, a DNA molecule corresponds to an $\ell$-length sequence over an alphabet of size 4, and the information is written onto $m$ DNA molecules stored in an unordered way. The storage capacity results in \cite{Heckel17} were extended to noisy settings in \cite{Ilan19arXiv} where the channel adds noise and randomly permutes the $m$ DNA molecules used to store information. The capacity results are obtained under the scenario where the length, $\ell$, of each DNA molecule grows with $m$. Although the effective channel in \cite{Ilan19arXiv} is closely related to the bee-identification channel in Fig.~\ref{Fig:BeeChannel}, we note that the fundamental problem in \cite{Ilan19arXiv} is to quantify the data storage capacity, while the main issue in the bee-identification problem is the identification of the row-permutation induced by the channel. 

Data communication over permutation channels with impairments was analyzed in \cite{Mladen18_Multisets}. The authors of \cite{Mladen18_Multisets} presented bounds on the size of optimal codes over a finite input alphabet, when the channel randomly permutes the letters of the input sequence in addition to causing impairments such as insertions, deletions, and substitutions. The effective channel for the bee-identification problem (see Fig.~\ref{Fig:BeeChannel}) differs from the communication channel in \cite{Mladen18_Multisets} in two aspects: (i) The input to the channel in the bee-identification problem is the entire codebook, not just a codeword belonging to the codebook. (ii)~The channel in Fig.~\ref{Fig:BeeChannel} only permutes the rows of the codebook, but does not permute the letters within a row.

\subsection{Bee-Identification Problem Formulation} \label{Sec:ProbFormulation}
The channel output is a row-permuted and noisy version of the codebook. If $\pi$ denotes a given permutation of $m$-letters, then the channel first permutes the $m$ rows of codebook $C$, based on $\pi$, to produce $C_{\pi}$ (see Fig.~\ref{Fig:BeeChannel}). Therefore, if $j=\pi(i)$ and the $i$-th row of codebook $C$ is denoted $\bfc_i = [c_{i,1}~ c_{i,2}~ \cdots~ c_{i,n}]$, then the $j$-th row of $C_{\pi}$ is equal to $\bfc_i$. The channel then applies noise on the permuted codebook $C_{\pi}$ to produce $\tilde{C}_{\pi}$, where noise is modeled by a BSC with crossover probability $p$, denoted BSC($p$), with $0 < p < 0.5$. If $j=\pi(i)$, and $\tilde{\bfc}_{\pi(i)}$ denotes the $j$-th row of $\tilde{C}_{\pi}$, then
\begin{align}
\Pr\{\tilde{\bfc}_{\pi(i)} | \bfc_i, \pi \} &= p^{d_i} (1-p)^{n-{d_i}} ,~~1 \le i \le m, \nonumber \\
\Pr\left\{\tilde{C}_{\pi} | C, \pi \right\} &= \prod_{i=1}^{m} \Pr\{\tilde{\bfc}_{\pi(i)} | \bfc_i, \pi \} = \prod_{i=1}^{m} p^{d_i} (1-p)^{n-{d_i}},  \label{eq:output_distribution}
\end{align}
where ${d_i} \triangleq \mathrm{d_H}(\tilde{\bfc}_{\pi(i)},\bfc_i)$ denotes the Hamming distance between vectors $\tilde{\bfc}_{\pi(i)}$ and $\bfc_i$. Let $\mathcal{M} \triangleq \{1,2,\ldots,m\}$, and let the decoder correspond to a function $\phi$ which takes $\tilde{C}_{\pi}$ as an input and produces a map $\nu : \mathcal{M} \to \mathcal{M}$ where $\nu(k)$ corresponds to the index of the transmitted codeword which produced the received word $\tilde{\bfc}_k$, for $1\le k \le m$. In effect, the bee-identification problem is that the decoder has to recover the row-permutation $\pi$ introduced by the channel, by using the knowledge of codebook $C$ and the channel output $\tilde{C}_{\pi}$.

\subsection{Bee-Identification Error Exponent} \label{Sec:IntroBeeExponent}
The indicator for the bee-identification error is defined as
\begin{equation*}
\mathcal{D}\left(\phi(\tilde{C}_{\pi}),\pi^{-1}\right)  = \mathcal{D}\left(\nu,\pi^{-1}\right) \triangleq \begin{cases}
1, ~~~\mbox{if } \nu\neq \pi^{-1}, \\
0, ~~~\mbox{if } \nu=\pi^{-1} .
\end{cases}
\end{equation*}
For a given codebook $C$ and decoding function $\phi$, the expected bee-identification error probability over the BSC($p$) is
\begin{equation}
D(C,p,\phi) \triangleq \mathbb{E}_{\pi} \left[  \mathbb{E}\left[\mathcal{D}\left(\phi(\tilde{C}_{\pi}),\pi^{-1}\right)\right] \right], \label{eq:Def_D}
\end{equation}
where the inner expectation is over the distribution of $\tilde{C}_{\pi}$ given $C$ and $\pi$ (see \eqref{eq:output_distribution}), and the outer expectation is over a uniform distribution of $\pi$ over all $m$-letter permutations. Note that \eqref{eq:Def_D} can be equivalently expressed as
\begin{equation}
D(C,p,\phi) =\Pr\Big\{\phi(\tilde{C}_{\pi}) \neq \pi^{-1}\Big\} =\Pr\left\{\nu \neq \pi^{-1}\right\}. \label{eq:Def_D_v2}
\end{equation}

For a given $R>0$, let the number of barcodes $m$ scale exponentially with blocklength $n$ as $m = 2^{nR}$. Now, for given values of $n$ and $R$, define the minimum expected bee-identification error probability as
\begin{equation}
\underline{D}(n,R,p) \triangleq \min_{C,\phi} D(C,p,\phi) , \label{eq:Def_min_D}
\end{equation}
where the minimum is over all codebooks $C$ of size $2^{nR} \times n$, and all decoding functions $\phi$. 

Define, $E_{\underline{D}}(R,p)$, the exponent corresponding to the minimum expected bee-identification error probability, as 
\begin{equation}
E_{\underline{D}}(R,p) = \liminf_{n \to \infty} \frac{-1}{n}\log \underline{D}(n,R,p). \label{eq:distortion_exponent}
\end{equation}

We introduce some notation that is used in the rest of the paper. We will denote $f(n) \doteq g(n)$ when $\lim_{n\to\infty} n^{-1} \log \left(f(n)/g(n)\right) = 0$. Similarly, we write $f(n) \dotleq g(n)$ (respectively, $f(n) \dotgeq g(n)$) if  $\limsup_{n\to\infty} n^{-1} \log \left(f(n)/g(n)\right) \le 0$ (respectively, $\ge 0$).% Further, if $t(n)$ is a non-decreasing positive function of $n$, $f(n) \dotleq g(n)$, and $a(n) = f(n)^{t(n)}$, then by a slight abuse of notation, we write $a(n) \dotleq g(n)^{t(n)}$ to imply $\limsup_{n\to\infty} \frac{1}{n\, t(n)} \log \left(a(n)/g(n)^{t(n)}\right) \le 0$.

\subsection{Our Contributions}
The ``bee-identification problem'' is introduced and the corresponding bee-identification exponent $E_{\underline{D}}(R,p)$ is analyzed in this paper. In particular, we provide the following \emph{explicit} bounds on this exponent.
\begin{itemize}
	\item A lower bound on $E_{\underline{D}}(R,p)$ using a random code ensemble (RCE) with independent barcode decoding (Sec.~\ref{Sec:RCE_NaiveDecoding}) and joint barcode decoding (Sec.~\ref{Sec:RCE_ML_Decoding}).
	\item A lower bound on $E_{\underline{D}}(R,p)$ using typical random codes (TRC) with independent barcode decoding (Sec.~\ref{Sec:TRC_Independent}) and joint barcode decoding (Sec.~\ref{Sec:TRC_Joint}).
	\item An upper bound on $E_{\underline{D}}(R,p)$ which is applicable to all possible codebook designs (Sec.~\ref{Sec:UB_IE_Exponent}).
\end{itemize}
We show that joint decoding of barcodes provides a significantly better exponent compared to separate decoding followed by learning the permutation. For low rates, we prove that the lower bound obtained using TRC is strictly better than the corresponding bound obtained using RCE. Further, as the rate approaches zero, we prove that the upper bound meets the lower bound obtained using TRC with joint barcode decoding.

\section{Random Code Ensemble} \label{Sec:RCE}
In this section, we present lower bounds on $E_{\underline{D}}(R,p)$ using an RCE~\cite{Barg02}. Let $\sC(n,R)$ denote the set of all binary matrices with $m = 2^{nR}$ rows and $n$ columns. Assume that codebook $C$ is uniformly distributed over $\sC(n,R)$. It is immediate from the definition of $\underline{D}(n,R,p)$ \eqref{eq:Def_min_D} that
\begin{equation}
\underline{D}(n,R,p) \le \frac{1}{|\sC(n,R)|} \sum_{C \in \sC(n,R)} D(C,p,\phi) , \label{eq:Opt_D_LessThan_Avg_D}
\end{equation}
where the expression on the right denotes the average performance using RCE. We proceed by quantifying this expression when the decoding function $\phi$ corresponds to: (i) independent barcode decoding (Sec.~\ref{Sec:RCE_NaiveDecoding}), and (ii) joint barcode decoding (Sec.~\ref{Sec:RCE_ML_Decoding}). The main results in this section are as follows: we present explicit lower bounds on $E_{\underline{D}}(R,p)$ using independent barcode decoding (Thm.~\ref{thm:RCE_ID_ErrExp_LB}) and joint barcode decoding (Thm.~\ref{thm:RCE_Exp_JointDecoding}). It is shown (Prop.~\ref{Prop:EtaIsBetter}) that the bee-identification exponent obtained using joint barcode decoding is strictly better than the corresponding exponent obtained with independent barcode decoding.

\subsection{Independent Decoding for Each Barcode} \label{Sec:RCE_NaiveDecoding}
Here, we analyze a na\"ive decoding strategy where each barcode is decoded independently. In this case, for $1 \le j \le m$, the decoder picks $\tilde{\bfc}_j$, the $j$-th row of $\tilde{C}_{\pi}$, and then decodes it to $\nu(j) = \argmin_k \mathrm{d_H}(\tilde{\bfc}_{j},\bfc_k)$. If there is more than one codeword at the same minimum Hamming distance from $\tilde{\bfc}_j$, then any one of the corresponding codeword indices is chosen at random. From \eqref{eq:Def_D_v2} and the union bound, we have
\begin{equation}
D(C,p,\phi) \le \sum_{j=1}^{m}\Pr\left\{\nu(j) \neq \pi^{-1}(j)\right\} . \label{eq:UnionBound_D_RC}
\end{equation}

Combining \eqref{eq:Opt_D_LessThan_Avg_D} and \eqref{eq:UnionBound_D_RC}, we get
\begin{equation}
\underline{D}(n,R,p) \le \sum_{j=1}^{m} \left(\sum_{C \in \sC(n,R)} \frac{\Pr\left\{\nu(j) \neq \pi^{-1}(j)\right\}}{|\sC(n,R)|} \right). \label{eq:RCE_ID_ErrProbUB}
\end{equation}
Now define
\begin{equation}
P(n,R,p) \triangleq \frac{1}{|\sC(n,R)|} \sum_{C \in \sC(n,R)} \Pr\left\{\nu(j) \neq \pi^{-1}(j)\right\}. \label{eq:RCE_ErrProbOverBSC}
\end{equation}
Note that $P(n,R,p)$ is independent of index $j$ due to the averaging over the ensemble of codebooks uniformly distributed over $\sC(n,R)$. For $i = \pi^{-1}(j)$, the expression for $P(n,R,p)$ corresponds to the probability of error when the $i$-th codeword is transmitted over BSC($p$). From \eqref{eq:RCE_ID_ErrProbUB} and \eqref{eq:RCE_ErrProbOverBSC}, we get
\begin{equation}
\underline{D}(n,R,p) \le m P(n,R,p). \label{eq:UpperBound_OptD}
\end{equation}
The following theorem uses \eqref{eq:UpperBound_OptD} to present an explicit lower bound on $E_{\underline{D}}(R,p)$. 
\begin{theorem} \label{thm:RCE_ID_ErrExp_LB}
	We have
	\begin{equation}
	E_{\underline{D}}(R,p) \ge |R_0(p) - 2R|^+ , \label{eq:identification-error_exponent_LB_v2}
	\end{equation}
	where $|x|^+ \triangleq \max(0,x)$, and 
	\begin{equation}
	R_0(p) \triangleq 1 - \log\left(1+\sqrt{4p(1-p)}\right). \label{eq:Def_R0}
	\end{equation}
\end{theorem}
\begin{IEEEproof}
	It is well known that the random coding exponent over BSC($p$), defined as $E_{\mathrm{r}}(R,p) \triangleq \liminf_{n \to \infty} (1/n)\log\left(1/P(n,R,p)\right)$, is given by~\cite{GallagerBook68,Barg02}
	\begin{numcases}{E_{\mathrm{r}}(R,p) =}
	R_0(p) - R, &$0 \le R \le R_{\mathrm{cr}}(p)$  \label{eq:RC_Exp_v1} \\
	D(\delta_{\GV}(R) \| p), &$R_{\mathrm{cr}}(p) \le R \le 1-H(p)$ \nonumber \\
	0, &$R \ge 1-H(p)$, \nonumber
	\end{numcases}
	where $H(\cdot)$ denotes the binary entropy function, $\delta_{\GV}(R)$ is the Gilbert-Varshamov (GV) distance~\cite{Barg02} defined as the value of $\delta$ in the interval $[0,0.5]$ with $H(\delta) = 1-R$, and $R_{\mathrm{cr}}(p)$ is the critical rate given by $R_{\mathrm{cr}}(p) = 1 - H\left(\frac{\sqrt{p}}{\sqrt{p} + \sqrt{1-p}}\right)$, and 
	\begin{equation*}
	D(x\|y) \triangleq x\log\frac{x}{y} + (1-x)\log\frac{1-x}{1-y}. 
	\end{equation*}
	Using the fact that $m = 2^{nR}$, and combining \eqref{eq:distortion_exponent}, \eqref{eq:UpperBound_OptD}, and the definition of $E_{\mathrm{r}}(R,p)$, we get
	\begin{equation}
	E_{\underline{D}}(R,p) \ge |E_{\mathrm{r}}(R,p) - R|^+ . \label{eq:RCE_BeeErrExp_LB_v1}
	\end{equation}
	Now, using explicit numerical computation, it can be shown that $R_0(p) \le 2 R_{\mathrm{cr}}(p)$. The proof is complete by combining \eqref{eq:RC_Exp_v1}, \eqref{eq:RCE_BeeErrExp_LB_v1}, and noting that $|E_{\mathrm{r}}(R,p) - R|^+ = 0$ when $R \ge R_{\mathrm{cr}}(p)$ because $E_{\mathrm{r}}(R,p)$ is a decreasing function of $R$.
\end{IEEEproof}
The lower bound on $E_{\underline{D}}(R,p)$ given by \eqref{eq:identification-error_exponent_LB_v2} was obtained by applying a na\"ive decoding strategy where each barcode was decoded independently. In the next subsection, we analyze the bee-identification exponent using joint barcode decoding.

\subsection{Joint Decoding of Barcodes} \label{Sec:RCE_ML_Decoding}
Let $S_m$ denote the set of permutations of $\{1,\ldots,m\}$. For joint maximum likelihood (ML) decoding of barcodes, the decoding function $\phi$ takes the noisy row-permuted codebook $\tilde{C}_\pi$ as input, and produces permutation $\nu = \rho^{-1}$ as output, where $\rho = \argmin_{\sigma \in S_m} \mathrm{d_H}(\tilde{C}_\pi,C_\sigma)$, and $\mathrm{d_H}(\tilde{C}_\pi,C_\sigma) \triangleq |\{(i,j) : \tilde{C}_\pi(i,j) \neq C_\sigma(i,j), 1\le i \le m, 1 \le j \le n\}|$. We aim to provide bounds on $\Pr\{\nu \neq \pi^{-1}\} = \Pr\{\rho \neq \pi\}$.

For any two permutations $\pi_1, \pi_2 \in S_m$, the sets of distances $\{\mathrm{d_H}(\tilde{C}_{\pi_1},C_\sigma)\}_{\sigma \in S_m}$ and $\{\mathrm{d_H}(\tilde{C}_{\pi_2},C_\sigma)\}_{\sigma \in S_m}$ are equal. Therefore, the performance of the joint ML decoder is independent of the channel permutation $\pi$, and we assume, without loss of generality, that the \emph{permutation induced by the channel is the identity permutation, denoted $\pi_0$.} 

For a given codebook $C$ at the transmitter, let $\tilde{C}_{\pi_0}$ denote the received noisy codebook at the output of the effective channel, and for $\sigma \in S_m$ with $\sigma \neq \pi_0$, we define
\begin{equation*}
\Pr\{\pi_0 \to \sigma\} \triangleq \Pr\left\{\mathrm{d_H}(\tilde{C}_{\pi_0}, C_{\sigma}) \le \mathrm{d_H}(\tilde{C}_{\pi_0}, C_{\pi_0}) \right\},
\end{equation*}
where the event $\{\pi_0 \to \sigma\}$ is said to occur if $\mathrm{d_H}(\tilde{C}_{\pi_0}, C_{\sigma}) \le \mathrm{d_H}(\tilde{C}_{\pi_0}, C_{\pi_0})$. 
From \eqref{eq:Def_D_v2}, we have
\begin{align}
D(C,p,\phi) &= \Pr\left\{\bigcup_{\sigma \in S_m,\linebreak\sigma \neq \pi_0} \{\pi_0 \to \sigma\}\right\}, \nonumber \\
&\le \sum_{\sigma \in S_m,\linebreak\sigma \neq \pi_0} \Pr\{\pi_0 \to \sigma\}, \label{eq:D_RC_ML_UnionBound}
\end{align}
where \eqref{eq:D_RC_ML_UnionBound} follows from the union bound. Now define
\begin{equation}
P_{\RCE,\sigma} \triangleq \frac{1}{|\sC(n,R)|} \sum_{C \in \sC(n,R)} \Pr\{\pi_0 \to \sigma\}, \label{eq:Def_P_RCE_sigma}
\end{equation}
which denotes the probability of the event $\{\pi_0 \to \sigma\}$, averaged over the ensemble of random binary codebooks. Using \eqref{eq:Opt_D_LessThan_Avg_D}, \eqref{eq:D_RC_ML_UnionBound}, and \eqref{eq:Def_P_RCE_sigma}, we get
\begin{equation}
\underline{D}(n,R,p) \le \sum_{\sigma \in S_m, \sigma \neq \pi_0} P_{\RCE,\sigma}. \label{eq:RCE_Joint_D_opt_Bound}
\end{equation}

Now consider two codewords $\bfc_{\hat{\imath}}$, $\bfc_{\hat{\jmath}}$ at distance $d$ from each other. Given that $\bfc_{\hat{\imath}}$ is transmitted over BSC($p$), the probability that the Hamming distance of the received word from  $\bfc_{\hat{\jmath}}$ is not more than its distance from $\bfc_{\hat{\imath}}$ is~\cite{Barg02}
\begin{equation*}
\Pr\{\bfc_{\hat{\imath}} \to \bfc_{\hat{\jmath}}\} \le 2^{-d\, \alpha_p}, 
\end{equation*}
where 
\begin{equation}
\alpha_p \triangleq -\log \sqrt{4p(1-p)}. \label{eq:alpha_p_def}
\end{equation}
Therefore, for a given codebook $C = C_{\pi_0}$ and permutation $\sigma \in S_m$ with $\sigma \neq \pi_0$, if $d_{\sigma} \triangleq \mathrm{d_H}(C_{\pi_0},C_{\sigma}) $, then 
\begin{equation}
\Pr\{\pi_0 \to \sigma\} \le 2^{-d_{\sigma} \alpha_p}. \label{eq:Prob_Decode_Sigma}
\end{equation}  
In the following, we quantify $P_{\RCE,\sigma}$ for different $\sigma \in S_m$, via \eqref{eq:Def_P_RCE_sigma} and \eqref{eq:Prob_Decode_Sigma}.

\subsubsection{$\sigma$ is a transposition}
We first consider the case where $\sigma$ is a \emph{transposition}, i.e. a  permutation that interchanges only two indices. For indices $\hat{\imath}, \hat{\jmath}$, with $1 \le \hat{\imath} < \hat{\jmath} \le m$, the Hamming distance between codewords $\bfc_{\hat{\imath}}$ and $\bfc_{\hat{\jmath}}$ in a random codebook satisfies~\cite{Barg02}
\begin{equation} 
\Pr\left\{\mathrm{d_H}(\bfc_{\hat{\imath}},\bfc_{\hat{\jmath}}) =d \right\} \le 2^{-n(1-H(d/n))}. \label{eq:RCE_d_prob}
\end{equation}
When $\sigma = (\hat{\imath}~\hat{\jmath})$ is the permutation that only transposes indices $\hat{\imath}$ and $\hat{\jmath}$, then $\mathrm{d_H}\left(C_{\pi_0}, C_{(\hat{\imath}~\hat{\jmath})}\right) = 2d$ if and only if $\mathrm{d_H}(\bfc_{\hat{\imath}},\bfc_{\hat{\jmath}}) =d$. Thus, it follows from \eqref{eq:RCE_d_prob} that $\Pr\left\{\mathrm{d_H}\left(C_{\pi_0}, C_{(\hat{\imath}~\hat{\jmath})}\right) = 2d\right\} \le 2^{-n(1-H(d/n))}$. Further, when $\mathrm{d_H}\left(C_{\pi_0}, C_{(\hat{\imath}~\hat{\jmath})}\right) = 2d$, we have $\Pr\{\pi_0 \to (\hat{\imath}~\hat{\jmath})\} \le 2^{-2d\, \alpha_p}$. Therefore, the probability $P_{\RCE,(\hat{\imath}~\hat{\jmath})}$ can be characterized using \eqref{eq:Def_P_RCE_sigma}, \eqref{eq:Prob_Decode_Sigma}, and \eqref{eq:RCE_d_prob} as
\begin{equation}
P_{\RCE,(\hat{\imath}~\hat{\jmath})} \le \sum_{d = 0}^n 2^{-n\left(1-H(d/n) + 2(d/n) \alpha_p\right)}. \label{eq:P_RC_Transposition_v1}
\end{equation}
If $\delta = d/n$ is treated as a continuous variable, then the exponent $
E_2(\delta) \triangleq 1-H(\delta) +2\delta\alpha_p$ is a convex function with a unique minimum at $\delta = \hat{\delta}_p$ where 
\begin{equation}
\hat{\delta}_p \triangleq \frac{4p(1-p)}{1 + 4p(1-p)}. \label{eq:Def_delta_hat}
\end{equation}
Therefore, for $0 \le d \le n$, we have
\begin{equation*}
2^{-n\left(1-H(d/n) + 2(d/n) \alpha_p\right)} \le 2^{-n\left(1-H(\hat{\delta}_p) + 2(\hat{\delta}_p) \alpha_p\right)}.
\end{equation*}
Now, if we define $c_n \triangleq \left(\log (n+1)\right)/n$, then it follows from \eqref{eq:P_RC_Transposition_v1} that
\begin{equation}
P_{\RCE,\sigma} \le 2^{-n\left(1-H(\hat{\delta}_p) + 2(\hat{\delta}_p) \alpha_p - c_n\right)}. \label{eq:P_RC_Transposition_v1b}
\end{equation}
Further, we have $1-H(\hat{\delta}_p) + 2(\hat{\delta}_p) \alpha_p = R_1(p)$, where
\begin{equation}
R_1(p) \triangleq 1 - \log(1+4p(1-p)). \label{eq:Def_R1}
\end{equation}
Hence, it follows from \eqref{eq:P_RC_Transposition_v1b} and \eqref{eq:Def_R1} that
\begin{equation}
P_{\RCE,\sigma} \le 2^{-n\left(R_1(p) - c_n\right)}, \label{eq:P_RC_Transposition_v2}
\end{equation}
where $\sigma$ is a transposition.

\subsubsection{$\sigma$ is a product (composition) of disjoint transpositions} \label{Sec:ProdOfTranspositions}
We now consider the case where $\sigma = \sigma_1 \sigma_2$, where $\sigma_1$ and $\sigma_2$ are disjoint transpositions with $\sigma_1 = (i~j)$ and $\sigma_2 = (\hat{\imath}~\hat{\jmath})$. As the codewords in a random codebook are independent, then using \eqref{eq:RCE_d_prob}, we have $\Pr\left\{ \{\mathrm{d_H}(\bfc_i,\bfc_j) = d_1\} \cap \{\mathrm{d_H}(\bfc_{\hat{\imath}},\bfc_{\hat{\jmath}}) = d_2\} \right\} \le \prod_{i=1}^2 2^{-n(1-H(d_i/n))}$. Further, if $\mathrm{d_H}(\bfc_i,\bfc_j) = d_1$ and $\mathrm{d_H}(\bfc_{\hat{\imath}},\bfc_{\hat{\jmath}}) = d_2$, then $\mathrm{d_H}\left(C_{\pi_0}, C_{\sigma}\right) = 2(d_1+d_2)$, and $\Pr\{\pi_0 \to \sigma\} \le 2^{-2(d_1+d_2) \alpha_p}$. Therefore, if $\sigma$ is a product of two disjoint transpositions, then 
\begin{align*}
P_{\RCE,\sigma} &\le \sum_{d_1,d_2}  2^{-n\left( \sum_{i=1}^2 (1-H(d_i/n) + 2(d_i/n) \alpha_p) \right)}, \nonumber \\
&= \prod_{i=1}^2 \left( \sum_{d_i=0}^n 2^{-n\left(1-H(d_i/n) + 2(d_i/n) \alpha_p\right)} \right), \nonumber \\
&\le 2^{-2n\left(R_1(p) - c_n\right)}. 
\end{align*}
In general, when $\sigma$ is a product of $s$ disjoint transpositions, the above argument can be readily extended to show that
\begin{equation}
P_{\RCE,\sigma} \le 2^{-sn\left(R_1(p) - c_n\right)} . \label{eq:P_RC_sProdTransposition_v2}
\end{equation}
Now, define
\begin{equation*}
\lambda_p \triangleq \min\left\{\frac{2R_0(p)}{3} ,\frac{R_1(p)}{2} \right\},
\end{equation*}
where $R_0(p)$ and $R_1(p)$ are defined in \eqref{eq:Def_R0} and \eqref{eq:Def_R1}, respectively. As $2\lambda_p \le R_1(p)$, it follows from \eqref{eq:P_RC_sProdTransposition_v2} that
\begin{equation}
P_{\RCE,\sigma} \le 2^{-n2s\left(\lambda_p - c_n\right)}. \label{eq:P_RC_sProdTransposition_v3}
\end{equation}
We remark that when $\sigma$ is just a transposition, then from \eqref{eq:P_RC_Transposition_v2} we have $P_{\RCE,\sigma} \le 2^{-n\left(R_1(p) - c_n\right)} \le 2^{-n2\left(\lambda_p - c_n\right)}$, which is only a special case of \eqref{eq:P_RC_sProdTransposition_v3} with $s=1$.

\subsubsection{$\sigma$ is a $k$-cycle with $k > 2$}
Let $\sigma \in S_m$ be a $k$-cycle $(i_1~i_2~\cdots~i_k)$ where $i_{l+1} = \sigma(i_l)$ for $1 \le l \le k-1$, and $i_1 = \sigma(i_k)$. We will apply the following proposition towards characterizing $P_{\RCE,\sigma}$.
\begin{proposition} \label{prop:DistanceIndependence_RCE}
	Let $\mathbb{F}_{2^n}$ denote the space of all $n$-length binary vectors. Let $\bfc_1, \bfc_2, \ldots, \bfc_k$ be $k>2$ $\mathrm{i.i.d.}$ random vectors, uniformly distributed over $\mathbb{F}_{2^n}$, and let $d_1, d_2, \ldots, d_{k-1}$ be given non-negative integers. Then the following holds
	\begin{equation}
	\Pr\left\{ \bigcap_{i=1}^{k-1} \left\{\mathrm{d_H}(\bfc_i,\bfc_{i+1}) = d_i \right\}\right\} \le \prod_{i=1}^{k-1} 2^{-n(1-H(d_i/n))}. \label{eq:independence_of_distances_v1}
	\end{equation}
\end{proposition}
\begin{IEEEproof}
	See Appendix~\ref{app:DistanceIndependence_RCE}.
\end{IEEEproof}
For a given codebook $C$, if $\mathrm{d_H}(\bfc_{i_l},\bfc_{i_{l+1}}) = d_l$ for $1 \le l \le k-1$, and $\mathrm{d_H}(\bfc_{i_k},\bfc_{i_{1}}) = d_k$, then $\mathrm{d_H}(C_{\pi_0}, C_{\sigma}) = \sum_{l=1}^k d_l$, and we have
\begin{equation}
\Pr\{\pi_0 \to \sigma\} \le 2^{-\left(\sum_{l=1}^k d_l\right) \alpha_p} . \label{eq:RCE_ErrProb_kcycle}
\end{equation}
Further, if codebook $C$ is uniformly distributed over $\sC(n,R)$,
\begin{align}
\Pr\bigg\{ \Big(\bigcap_{l=1}^{k-1} &\left\{\mathrm{d_H}(\bfc_{i_l},\bfc_{i_{l+1}}) = d_l\right\}\Big) \bigcap \left\{\mathrm{d_H}(\bfc_{i_k},\bfc_{i_{1}}) = d_k \right\}  \bigg\} \nonumber \\
&\le 2^{-n \left(\sum_{l=1}^{k-1} (1 - H(d_l/n)) \right)}, \label{eq:independence_of_distances_v4}
\end{align}
where \eqref{eq:independence_of_distances_v4} follows from \eqref{eq:independence_of_distances_v1}. Combining \eqref{eq:RCE_ErrProb_kcycle} and \eqref{eq:independence_of_distances_v4},
\begin{align}
P_{\RCE,\sigma} &\le \sum_{\substack{0 \le d_l \le n,\\1 \le l \le k}} 2^{-n \left( (\sum_{l=1}^k (d_l/n)\alpha_p) + (\sum_{l=1}^{k-1} (1 - H(d_l/n))) \right)}, \nonumber \\
&= \sum_{d_k=0}^n 2^{-d_k \alpha_p} \left(\prod_{l=1}^{k-1} \sum_{d_l = 0}^n 2^{-n \left(1 - H(d_l/n) + (d_l/n)\alpha_p \right)}\right) \nonumber \\
&\le \prod_{l=1}^{k-1} \sum_{d_l = 0}^n 2^{-n \left(1 - H(d_l/n) + (d_l/n)\alpha_p \right)}. \label{eq:P_RC_kcycle_v1}
\end{align}
If $\delta = d_l/n$ is treated as a continuous variable, then the exponent $E_1(\delta) \triangleq 1-H(\delta) + \delta\alpha_p$ is a convex function with a unique minimum at $\delta = \tilde{\delta}_p$, where
\begin{equation}
\tilde{\delta}_p \triangleq \frac{\sqrt{4p(1-p)}}{1 + \sqrt{4p(1-p)}}. \label{eq:Def_delta_tilde}
\end{equation}
We have 
\begin{equation*}
E_1(\tilde{\delta}_p) = 1 - \log(1+\sqrt{4p(1-p)}) = R_0(p), 
\end{equation*}
and therefore
\begin{equation}
\sum_{d_l = 0}^n 2^{-n \left(1 - H(d_l/n) + (d_l/n)\alpha_p \right)} \le 2^{-n \left(R_0(p) - c_n\right)}, \label{eq:P_RC_kcycle_v1b} 
\end{equation}
where $c_n = \left(\log(n+1)\right)/n$. Combining \eqref{eq:P_RC_kcycle_v1} and \eqref{eq:P_RC_kcycle_v1b},
\begin{equation}
P_{\RCE,\sigma} \le 2^{-n (k-1) \left(R_0(p) - c_n\right)}. \label{eq:P_RC_kcycle_v2}
\end{equation}
As $2k/3 \le k-1$ for $k>2$, we have $k \lambda_p \le 2kR_0(p)/3 \le (k-1)R_0(p)$, and it follows from \eqref{eq:P_RC_kcycle_v2} that
\begin{equation}
P_{\RCE,\sigma} \le 2^{-n k \left(\lambda_p - c_n\right)}. \label{eq:P_RC_kcycle_v3}
\end{equation}
The above equation has been derived for the case where $\sigma$ is a $k$-cycle with $k>2$. However, a transposition is just a $k$-cycle with $k=2$, and from the remark following \eqref{eq:P_RC_sProdTransposition_v3}, it follows that \eqref{eq:P_RC_kcycle_v3} holds even for $k=2$.

\subsubsection{General $\sigma \in S_m$ with $\sigma \neq \pi_0$}
It is well known that any permutation $\sigma \neq \pi_0$ can be written as a product (composition) of $t$ disjoint cycles, for $t \ge 1$~\cite{HersteinBook75}. Consider a given $\sigma$ which is a product of $t$ disjoint cycles of length $k_1, \ldots, k_t$, respectively, where $k_i \ge 2$ for $1 \le i \le t$. Then, we can extend the result in \eqref{eq:P_RC_kcycle_v3} to obtain
\begin{equation}
P_{\RCE,\sigma} \le 2^{-n \left(\sum_{i=1}^t k_i \right)\left(\lambda_p - c_n\right)}. \label{eq:P_RC_general_sigma}
\end{equation}

\subsubsection{Putting it all together}
For $1 \le j \le m$, if we define
\begin{align}
&\Sigma_j \triangleq \left\{\sigma \in S_m: |\{i:\sigma(i)\neq i, 1\le i \le m\}|=j\right\}, \label{eq:Sigma_def} \\
&P_{\RCE,\Sigma_j} \triangleq \sum_{\sigma \in \Sigma_j} P_{\RCE,\sigma}~, \label{eq:PSigma_def}
\end{align}
then \eqref{eq:RCE_Joint_D_opt_Bound} can be equivalently expressed as
\begin{equation}
\underline{D}(n,R,p) \le \sum_{j=2}^m P_{\RCE,\Sigma_j}. \label{eq:RCE_Joint_D_opt_Bound_v2}
\end{equation}
Note that the set $\Sigma_1$ is empty, as the Hamming distance between two distinct permutations is at least two. The set $\Sigma_2$ consists of all transpositions and $|\Sigma_2| = \binom{m}{2} \le 2^{n(2R)}$. For all $\sigma \in \Sigma_2$, the value of $P_{\RCE,\sigma}$ is given by \eqref{eq:P_RC_Transposition_v2}, and combining this with \eqref{eq:PSigma_def}, we get
\begin{equation}
P_{\RCE,\Sigma_2} \le 2^{-n(R_1(p) - c_n - 2R)}. \label{eq:P_RC_Sigma2}
\end{equation}
For a given $j > 2$, if $\sigma \in \Sigma_j$, then from \eqref{eq:P_RC_general_sigma} it follows that $P_{\RCE,\sigma} \le 2^{-n j \left(\lambda_p - c_n\right)}$. For $j \ge 2$, the size of the set $\Sigma_j$ satisfies $|\Sigma_j| < \prod_{i=0}^{j-1} (m-i) < 2^{njR}$. If we define $$\beta_n \triangleq 2^{-n (\lambda_p - c_n - R)} ,$$then we have $P_{\RCE,\Sigma_j} \le \beta_n^j$. Now, if $R < \lambda_p$, then because $c_n = o(1)$, there exists $N$ such that for $n \ge N$, we have $R < \lambda_p - c_n$ and hence $\beta_n < 1$. Therefore, for $n \ge N$, 
\begin{equation}
\sum_{j=3}^m P_{\RCE,\Sigma_j} \le \sum_{j=3}^{m} \beta_n^j \le \frac{\beta_n^3}{1-\beta_n}. \label{eq:P_RC_Sigma_j_Sum_v0}
\end{equation}
As $\beta_n \to 0$  and $c_n \to 0$ when $n \to \infty$, it follows from \eqref{eq:P_RC_Sigma_j_Sum_v0} that
\begin{equation}
\sum_{j=3}^m P_{\RCE,\Sigma_j} \le \frac{\beta_n^3}{1-\beta_n} \doteq \beta_n^3 \doteq 2^{-3n(\lambda_p - R)}. \label{eq:P_RC_Sigma_j_Sum}
\end{equation}
Combining \eqref{eq:RCE_Joint_D_opt_Bound_v2}, \eqref{eq:P_RC_Sigma2}, and \eqref{eq:P_RC_Sigma_j_Sum}, for $R < \lambda_p$,
\begin{equation}
\underline{D}(n,R,p) \dotleq 2^{-n(R_1(p) - 2R)} + 2^{-n (3 \lambda_p - 3R)}. \label{eq:UpperBound_min_D_JointDecoding}
\end{equation}
Comparing \eqref{eq:RCE_Joint_D_opt_Bound} with \eqref{eq:UpperBound_min_D_JointDecoding}, we observe that the error probability $\underline{D}(n,R,p)$ is dominated by $P_{\RCE,\sigma}$ terms for $\sigma$ corresponding to $k$-cycles with $k=2$ and $k=3$. The next theorem presents an explicit lower bound for $E_{\underline{D}}(R,p)$ when the decoder jointly decodes all the barcodes using a maximum likelihood approach.	
\begin{theorem} \label{thm:RCE_Exp_JointDecoding}
	We have
	\begin{equation}
	E_{\underline{D}}(R,p) \ge |\eta_p(R)|^+ , \label{eq:identification-error_exponent_LB_JointDecoding}
	\end{equation}
	where $\eta_p(R) \triangleq \min\left\{R_1(p) - 2R,\, 2R_0(p) - 3R\right\}$.
\end{theorem}
\begin{IEEEproof}
	If $R < \lambda_p$, then $R_1(p) \ge 2\lambda_p > 2R$. Therefore, from \eqref{eq:UpperBound_min_D_JointDecoding} it follows that if $R < \lambda_p$, then $E_{\underline{D}}(R,p)$ is lower bounded by $\min\left\{R_1(p)-2R, \ 3\lambda_p - 3R\right\} = \eta_p(R)$. Further, note that $\eta_p(R) > 0$ if and only if $R < \lambda_p$.
\end{IEEEproof}
The following proposition shows that the lower bound~\eqref{eq:identification-error_exponent_LB_JointDecoding} (obtained using joint decoding of barcodes) is \emph{strictly better} than the bound given by~\eqref{eq:identification-error_exponent_LB_v2} (obtained with independent decoding of barcodes) in the interval where it is positive.
\begin{proposition} \label{Prop:EtaIsBetter}
When $R_0(p) > 2R$ and $0 < p < 0.5$, then we have the strict inequality
\begin{equation*}
\eta_p(R) > R_0(p) - 2R. \label{eq:EtaIsBetter}
\end{equation*}
\end{proposition}
\begin{IEEEproof}
When $0 < p < 0.5$, we have $0 < 4p(1-p) < \sqrt{4p(1-p)} < 1$, and hence $R_1(p) > R_0(p)$. If $R_0(p) > 2R$, then $2R_0(p) - 3R = 2(R_0(p)-2R) + R > R_0(p)-2R$. The proof is complete by combining these observations with the definition of $\eta_p(R)$.
\end{IEEEproof}
Note that $|\eta_p(R)|^+ = 0$ for $R \ge 0.5$, because in this case $\eta_p(R) \le R_1(p)-2R \le R_1(p) - 1 \le 0$. In the following section, we present improved lower bounds on $E_{\underline{D}}(R,p)$ by analyzing \emph{typical} random codebooks.

\section{Typical Random Code}
TRCs are known, in general, to provide higher error exponents than RCE over a BSC~\cite{Barg02,Merhav18_TRC}. Roughly speaking, TRCs are characterized by the property that their relative minimum distance is at least $\delta_{\GV}(2R)$. Formally, for $0 \le R < 0.5$, $0 < \epsilon <  \delta_{\GV}(2R)$, and indices $1 \le \hat{\imath} < \hat{\jmath} \le m=2^{nR}$, the Hamming distance between codewords $\bfc_{\hat{\imath}}$ and $\bfc_{\hat{\jmath}}$ in a TRC satisfies~\cite{Barg02}
\begin{equation} \label{eq:TRC_d_prob}
\Pr\left\{\mathrm{d_H}(\bfc_{\hat{\imath}},\bfc_{\hat{\jmath}}) =d \right\} \begin{cases}
\le 2^{-n(1-H(\delta))}, & |\frac{1}{2} - \delta| \le \frac{1}{2} - \overline{\delta} \\
=0, & |\frac{1}{2} - \delta| \ge \frac{1}{2} - \underline{\delta},
\end{cases}
\end{equation}
where $\delta = d/n$, $\overline{\delta} \triangleq \delta_{\GV}(2R) + \epsilon$, and $\underline{\delta} \triangleq \delta_{\GV}(2R) - \epsilon$. 

Let $\sC_{\TRC}(n,R)$ denote the set of all codebooks of size $2^{nR} \times n$, with the property that the Hamming distance between a pair of codewords $\bfc_i$ and $\bfc_j$ satisfies the relation $n\underline{\delta} < \mathrm{d_H}(\bfc_i,\bfc_j) < n(1-\underline{\delta})$ for all $i \neq j$. Note that if codebook $C$ is uniformly distributed over $\sC_{\TRC}(n,R)$, then the Hamming distance between a pair of distinct codewords satisfies~\eqref{eq:TRC_d_prob}. It is immediate from \eqref{eq:Def_min_D} that
\begin{equation}
\underline{D}(n,R,p) \le \frac{1}{|\sC_{\TRC}(n,R)|} \sum_{C \in \sC_{\TRC}(n,R)} D(C,p,\phi) , \label{eq:Opt_D_LessThan_TRC_D}
\end{equation}
where the expression on the right denotes the average performance using TRCs. 

In this section we provide lower bounds on the bee-identification exponent $E_{\underline{D}}(R,p)$ using TRCs. The case where each barcode is decoded independently is analyzed in Sec.~\ref{Sec:TRC_Independent} while joint barcode decoding is analyzed in Sec.~\ref{Sec:TRC_Joint}. It is shown that these lower bounds on $E_{\underline{D}}(R,p)$ using TRCs outperform the corresponding bounds for RCEs when the rate is smaller than a certain threshold. 

\subsection{Independent Decoding of Barcodes} \label{Sec:TRC_Independent}
With independent barcode decoding, the decoder picks $\tilde{\bfc}_j$, the $j$-th row of $\tilde{C}_{\pi}$, and then assigns $\nu(j) = \argmin_k \mathrm{d_H}(\tilde{\bfc}_{j},\bfc_k)$, for $1 \le j \le m$. From the union bound, we have $D(C,p,\phi) \le \sum_{j=1}^{m}\Pr\left\{\nu(j) \neq \pi^{-1}(j)\right\}$, and using \eqref{eq:Opt_D_LessThan_TRC_D} we get
\begin{equation}
\underline{D}(n,R,p) \le \sum_{j=1}^{m} \left(\sum_{C \in \sC_{\TRC}(n,R)} \frac{\Pr\left\{\nu(j) \neq \pi^{-1}(j)\right\}}{|\sC_{\TRC}(n,R)|} \right). \label{eq:TRC_ID_ProbErrorUB}
\end{equation}
Let $P_{\TRC}(n,R,p) \triangleq \sum_{C \in \sC_{\TRC}(n,R)} \frac{\Pr\left\{\nu(j) \neq \pi^{-1}(j)\right\}}{|\sC_{\TRC}(n,R)|}$. Note that $P_{\TRC}(n,R,p)$ is independent of the index $j$ due to the symmetry resulting from averaging over codebooks uniformly distributed over $\sC_{\TRC}(n,R)$. For $i = \pi^{-1}(j)$, the expression for $P_{\TRC}(n,R,p)$ corresponds to the probability of error when the $i$-th codeword is transmitted. From \eqref{eq:TRC_ID_ProbErrorUB}, we get
\begin{equation}
\underline{D}(n,R,p) \le m P_{\TRC}(n,R,p). \label{eq:TRC_UpperBound_OptD}
\end{equation}
The following theorem uses \eqref{eq:TRC_UpperBound_OptD} to present an explicit lower bound on $E_{\underline{D}}(R,p)$ when the rate is smaller than a certain threshold. 
\begin{theorem} \label{thm:TRC_ID_ErrExp_LB}
We have 
\begin{equation}
E_{\underline{D}}(R,p) \ge \alpha_p \delta_{\GV}(2R), ~~0 \le R < R_{\TRC}(p), \label{eq:TRC_ErrExp_LB}
\end{equation}
where $\alpha_p$ is defined in \eqref{eq:alpha_p_def}, and 
\begin{equation}
R_{\TRC}(p) \triangleq 0.5 \left(1 - H\left(\frac{\sqrt{4p(1-p)}}{1+\sqrt{4p(1-p)}}\right)\right).  \label{eq:Def_R_T}
\end{equation}
\end{theorem}
\begin{IEEEproof}
It is known that for $0 \le R < R_{\TRC}(p) \le 0.5$, the error exponent using a TRC over BSC($p$), defined as $E_{\TRC}(R,p) \triangleq \liminf_{n \to \infty} (1/n)\log \left(1/P_{\TRC}(n,R,p)\right)$, is given by~\cite{Barg02}
\begin{equation}
E_{\TRC}(R,p) = \alpha_p \delta_{\GV}(2R) + R. \label{eq:E_TRC_value}
\end{equation}
Using the fact that $m = 2^{nR}$, and combining \eqref{eq:distortion_exponent}, \eqref{eq:TRC_UpperBound_OptD}, with the definition of $E_{\TRC}(R,p)$, we get
\begin{equation}
E_{\underline{D}}(R,p) \ge |E_{\TRC}(R,p) - R|^+ . \label{eq:TRC_ErrExp_LB_v0}
\end{equation}
The proof is completed by applying \eqref{eq:E_TRC_value} in \eqref{eq:TRC_ErrExp_LB_v0}.
\end{IEEEproof}

It is well known that $E_{\TRC}(R,p) > E_{\mathrm{r}}(R,p)$ for $0 \le R < R_{\TRC}(p)$~\cite{Barg02}. This implies that the lower bound on $E_{\underline{D}}(R,p)$ for TRC given by \eqref{eq:TRC_ErrExp_LB} is \emph{strictly better} than the corresponding bound for RCE given by \eqref{eq:identification-error_exponent_LB_v2} when $0 \le R < R_{\TRC}(p)$. The next subsection provides a more refined bound on $E_{\underline{D}}(R,p)$ by analyzing joint decoding of barcodes using TRCs.

\subsection{Joint Decoding of Barcodes} \label{Sec:TRC_Joint}
With joint barcode decoding, the decoder takes the noisy row-permuted codebook $\tilde{C}_\pi$ as input, and produces the permutation $\nu = \rho^{-1}$ as output, where $\rho = \argmin_{\sigma \in S_m} \mathrm{d_H}(\tilde{C}_\pi,C_\sigma)$. As in Sec.~\ref{Sec:RCE_ML_Decoding}, we assume, without loss of generality, that the permutation induced by the channel is the identity permutation $\pi_0$. For a given codebook $C$, we have $D(C,p,\phi) \le \sum_{\sigma \in S_m, \sigma \neq \pi_0} \Pr\{\pi_0 \to \sigma\}$. If we define 
\begin{equation}
P_{\TRC,\sigma} \triangleq \mathbb{E}\left[\Pr\{\pi_0 \to \sigma\}\right], \label{eq:Def_P_TRC_sigma}
\end{equation}
where the expectation is over a uniform distribution of codebook over $\sC_{\TRC}(n,R)$, then we have
\begin{align}
\underline{D}(n,R,p) &\le \mathbb{E}\left[D(C,p,\phi)\right], \nonumber \\
&\le \sum_{\sigma \in S_m, \sigma \neq \pi_0} P_{\TRC,\sigma}. \label{eq:TRC_Opt_D_UnionBound}
\end{align}
In the following, we quantify $P_{\TRC,\sigma}$ for different $\sigma \in S_m$, in order to bound $\underline{D}(n,R,p)$ via \eqref{eq:TRC_Opt_D_UnionBound}.

\subsubsection{$\sigma$ is a transposition}
If $\sigma = (\hat{\imath}~\hat{\jmath})$ is the permutation that only transposes indices $\hat{\imath}$ and $\hat{\jmath}$, and $\mathrm{d_H}(\bfc_{\hat{\imath}},\bfc_{\hat{\jmath}}) =d$, then $\mathrm{d_H}\left(C_{\pi_0}, C_{(\hat{\imath}~\hat{\jmath})}\right) = 2d$, and we have 
\begin{equation}
\Pr\{\pi_0 \to (\hat{\imath}~\hat{\jmath})\} \le 2^{-2d \alpha_p}. \label{eq:TRC_TranspositionErrProb}
\end{equation}
When $C$ is uniformly distributed $\sC_{\TRC}(n,R)$ and $n \underline{\delta} \le d \le n(1-\underline{\delta})$, then
\begin{align}
\Pr\left\{\mathrm{d_H}\left(C_{\pi_0}, C_{(\hat{\imath}~\hat{\jmath})}\right) = 2d\right\} &= \Pr\left\{\mathrm{d_H}(\bfc_{\hat{\imath}},\bfc_{\hat{\jmath}}) =d \right\}, \nonumber \\
&\le 2^{-n(1-H(d/n))}, \label{eq:Transposition_2d_prob}
\end{align}
where \eqref{eq:Transposition_2d_prob} follows from \eqref{eq:TRC_d_prob}. Combining \eqref{eq:Def_P_TRC_sigma}, \eqref{eq:TRC_TranspositionErrProb}, and \eqref{eq:Transposition_2d_prob}, we get
\begin{equation}
P_{\TRC,(\hat{\imath}~\hat{\jmath})} \le \sum_{d = n \underline{\delta}}^{n(1-\underline{\delta})} 2^{-n(1-H(d/n)+2 (d/n) \alpha_p)}. \label{eq:P_TRC_2cycle}
\end{equation}
If $\delta = d/n$ is treated as a continuous variable, then the exponent $ E_2(\delta) = 1 - H(\delta) + 2 \delta \alpha_p$ is a convex function of $\delta$ with a unique minimum at $\hat{\delta}_p$ defined in \eqref{eq:Def_delta_hat}. If we define
\begin{equation}
\hat{R}_p \triangleq 0.5(1-H(\hat{\delta}_p)), \label{eq:Def_R_hat_p}
\end{equation}
then for $0 \le R < \hat{R}_p$, we have
\begin{equation*}
\delta_{\GV}(2R) > \delta_{\GV}(2 \hat{R}_p) = \hat{\delta}_p.
\end{equation*}
The exponent $E_2(\delta)$ increases monotonically in $\delta$ for $\delta \ge \hat{\delta}_p$. Therefore, if $0 \le R < \hat{R}_p$ and $\epsilon < \delta_{\GV}(2R) - \hat{\delta}_p$, the exponent in \eqref{eq:P_TRC_2cycle} is minimized for $d = n \underline{\delta}$, and we have
\begin{equation}
P_{\TRC,(\hat{\imath}~\hat{\jmath})} \le 
2^{-n(1 - H(\underline{\delta}) + 2 \underline{\delta} \alpha_p - c_n)}, ~~~ 0 \le R < \hat{R}_p, \label{eq:P_TRC_2cycle_v2}
\end{equation}
where $c_n = \left(\log(n+1)\right)/n$.

\subsubsection{$\sigma$ is a $k$-cycle}
We now consider the case where $\sigma$ is a $k$-cycle with $k \ge 3$. We will apply the following proposition towards characterizing $P_{\TRC,\sigma}$.
\begin{proposition} \label{prop:WeakIndependence_TRC}
Let $\bfc_{i_1}, \bfc_{i_2}, \ldots, \bfc_{i_k}$ be $k$ distinct rows in codebook $C$, and let $d_l$ satisfy $n \underline{\delta} \leq d_l \leq n \left(1-\underline{\delta}\right)$ for $1 \le l \le k-1$. Let $Q_{\TRC}\left\{ \bigcap_{l=1}^{k-1} \left\{\mathrm{d_H}(\bfc_{i_l},\bfc_{i_{l+1}}) = d_l\right\} \right\}$ denote the probability $\Pr\left\{ \bigcap_{l=1}^{k-1} \left\{\mathrm{d_H}(\bfc_{i_l},\bfc_{i_{l+1}}) = d_l\right\} \right\}$ when $C$ is uniformly distributed over $\sC_{\TRC}(n,R)$. Then, we have
	\begin{equation}
	Q_{\TRC}\left\{ \bigcap_{l=1}^{k-1} \left\{\mathrm{d_H}(\bfc_{i_l},\bfc_{i_{l+1}}) = d_l\right\} \right\} \le \frac{1}{\alpha_n} \prod_{l=1}^{k-1} 2^{-n(1 - H(d_l/n))}, \label{eq:WeakIndependence_TRC}
	\end{equation}
	where
	\begin{equation}
	\alpha_n \triangleq \sum_{(\gamma_1, \gamma_2, \ldots, \gamma_m) \in \sC_{\TRC}(n,R)} Q_{\RCE}\left\{ \bigcap_{i=1}^{m} \{\bfc_i = \gamma_i\}\right\},
	\end{equation}
	and $Q_{\RCE}\left\{ \bigcap_{i=1}^{m} \{\bfc_i = \gamma_i\}\right\}$ denotes the probability $\Pr\left\{ \bigcap_{i=1}^{m} \{\bfc_i = \gamma_i\}\right\}$ when $C$ is uniformly distributed over $\sC(n,R)$.	
\end{proposition}
\begin{IEEEproof}
	See Appendix~\ref{app:WeakIndependence_TRC}.
\end{IEEEproof}
Now, given that $\sigma = (i_1~i_2~\cdots~i_k)$ and $\mathrm{d_H}(\bfc_{i_l},\bfc_{i_{l+1}}) = d_l$ for $1 \le l \le k-1$, and $\mathrm{d_H}(\bfc_{i_k},\bfc_{i_{1}}) = d_k$, we have $\mathrm{d_H}(C_{\pi_0}, C_{\sigma}) = \sum_{l=1}^k d_l$, and therefore
\begin{equation}
\Pr\{\pi_0 \to \sigma\} \le 2^{-\left(\sum_{l=1}^k d_l\right) \alpha_p} . \label{eq:ErrProb_kcycle}
\end{equation}
If $d_0 \triangleq n \underline{\delta}$, then combining \eqref{eq:WeakIndependence_TRC} and \eqref{eq:ErrProb_kcycle}, we get
\begin{align}
P_{\TRC,\sigma} &\le \sum_{\substack{d_0 \le d_l \le n-d_0,\\1 \le l \le k}} \bigg( 2^{-n (\sum_{l=1}^k (d_l/n)\alpha_p)} \nonumber \\
	&~~~~~~~~~~~~~~~~~~~~~~~~~\times \frac{1}{\alpha_n}  2^{-n (\sum_{l=1}^{k-1} (1 - H(d_l/n)))} \bigg), \nonumber \\
&= \frac{1}{\alpha_n} \, \eta_k ~\prod_{l=1}^{k-1} \zeta_l , \label{eq:P_TRC_kcycle_v1}
\end{align}
where, for $1 \le l \le k-1$, we have
\begin{align}
\zeta_l &\triangleq \sum_{d_0 \le d_l \le n-d_0} 2^{-n \left(1 - H(d_l/n) + (d_l/n)\alpha_p \right)}, \mathrm{~and~} \label{eq:zeta_l_def} \\
\eta_k &\triangleq \sum_{d_0 \le d_k \le n-d_0} 2^{-d_k \alpha_p} \le 2^{-n \left(\underline{\delta} \alpha_p - c_n\right)}. \label{eq:eta_k_def}
\end{align}
The function $E_1(\delta) = 1 - H(\delta) + \delta \alpha_p$ is a convex function of $\delta$, and has a unique minimum that occurs at $\tilde{\delta}_p$ defined in \eqref{eq:Def_delta_tilde}. 
From \eqref{eq:Def_R_T} we observe that $R_{\TRC}(p) = 0.5(1-H(\tilde{\delta}_p))$. Thus, if $R < R_{\TRC}(p)$, then we have $\delta_{\GV}(2R) > \tilde{\delta}_p$. Further, $E_1(\delta)$ is an increasing function of  $\delta$ for $\delta \ge \tilde{\delta}_p$, and so if $R < R_{\TRC}(p)$ and $\epsilon < \delta_{\GV}(2R) - \tilde{\delta}_p$, the exponent in \eqref{eq:zeta_l_def} is minimized when $d_l = d_0 = n \underline{\delta}$. Thus, we have
\begin{equation}
\zeta_l \le 2^{-n (1 - H(\underline{\delta}) + \underline{\delta} \alpha_p - c_n)}, ~~0 \le R < R_{\TRC}(p). \label{eq:zeta_l_approx}
\end{equation}
Combining \eqref{eq:P_TRC_kcycle_v1}, \eqref{eq:eta_k_def}, and \eqref{eq:zeta_l_approx}, for $0 \le R < R_{\TRC}(p)$, 
\begin{equation}
P_{\TRC,\sigma} \le \frac{1}{\alpha_n} 2^{-n \left( (k-1)(1 - H(\underline{\delta})) + k (\underline{\delta} \alpha_p - c_n) \right)}, \label{eq:P_TRC_kcycle_v2}
\end{equation}
where $\sigma$ is a $k$-cycle with $k > 2$. As $k < 2(k-1)$ for $k > 2$, it follows from \eqref{eq:P_TRC_kcycle_v2} that
\begin{equation}
P_{\TRC,\sigma} \le \frac{1}{\alpha_n} 2^{-n k \left(0.5 (1 - H(\underline{\delta})) + \underline{\delta} \alpha_p - c_n \right)}, ~0 \le R < R_{\TRC}(p). \label{eq:P_TRC_kcycle_v3}
\end{equation}
Recall that $\hat{\delta}_p$ and $\hat{R}_p$ are given by \eqref{eq:Def_delta_hat} and \eqref{eq:Def_R_hat_p}, respectively. As $x/(1+x)$ is an increasing function of $x$, and $0 < p < 0.5$, it follows that $\hat{\delta}_p < \tilde{\delta}_p < 0.5$, which implies that $R_{\TRC}(p) < \hat{R}_p$. Note that a transposition is simply a $k$-cycle with $k=2$, and comparing \eqref{eq:P_TRC_2cycle_v2} with \eqref{eq:P_TRC_kcycle_v3} we observe that the relation given by \eqref{eq:P_TRC_kcycle_v3} holds even when $k=2$.

\subsubsection{$\sigma$ is a product (composition) of two disjoint cycles}
We now consider the case where $\sigma = \sigma_1 \sigma_2$, where $\sigma_1$ and $\sigma_2$ are disjoint cycles of length $k_1$ and $k_2$, respectively. Let $\sigma_1 = (i_1~i_2~\cdots~i_{k_1})$ and $\sigma_2 = (i_{k_1+1}~i_{k_1+1}~\cdots~i_{k_1+k_2})$. If $d_0 \le d_l \le n-d_0$ for $1 \le l \le k_1+k_2$, then a straightforward extension of Prop.~\ref{prop:WeakIndependence_TRC} shows that the probability $\Pr\Big\{ \bigcap_{l=1}^{k_1-1} \{\mathrm{d_H}(\bfc_{i_l},\bfc_{i_{l+1}}) = d_l\} \bigcap \left\{\mathrm{d_H}(\bfc_{i_{k_1}},\bfc_{i_{1}}) = d_{k_1} \right\} \\ ~~~~~~\bigcap_{l=k_1 + 1}^{k_1 + k_2 - 1} \left\{\mathrm{d_H}(\bfc_{i_l},\bfc_{i_{l+1}}) = d_l\right\} \\
~~~~~~\bigcap \left\{\mathrm{d_H}(\bfc_{i_{k_1+k_2}},\bfc_{i_{k_1 + 1}}) = d_{k_1 + k_2} \right\} \Big\}$ is upper bounded by
%\begin{align}
%&\Pr\Bigg\{\Big(\bigcap_{l=1}^{k_1-1} \{\mathrm{d_H}(\bfc_{i_l},\bfc_{i_{l+1}}) = d_l\}\Big) \bigcap \left\{\mathrm{d_H}(\bfc_{i_{k_1}},\bfc_{i_{1}}) = d_{k_1} \right\}   \nonumber \\
%&~~~~~~\bigcap_{l=k_1 + 1}^{k_1 + k_2 - 1} \left\{\mathrm{d_H}(\bfc_{i_l},\bfc_{i_{l+1}}) = d_l\right\} \nonumber \\
%&~~~~~~~~~~~~~~\bigcap \left\{\mathrm{d_H}(\bfc_{i_{k_1+k_2}},\bfc_{i_{k_1 + 1}}) = d_{k_1 + k_2} \right\} \Bigg\} \nonumber \\
%&\dotleq 2^{-n \left(\sum_{l=1}^{k_1-1} (1 - H(d_l/n)) \right)}\times 2^{-n \left(\sum_{l=k_1+1}^{k_1+k_2-1} (1 - H(d_l/n)) \right)} . \label{eq:WeakIndependence_v3}
%\end{align}
\begin{equation}
\frac{1}{\alpha_n} 2^{-n \left(\sum_{l=1}^{k_1-1} (1 - H(d_l/n)) \right)}\times 2^{-n \left(\sum_{l=k_1+1}^{k_1+k_2-1} (1 - H(d_l/n)) \right)} . \label{eq:WeakIndependence_v3}
\end{equation}
Further, for a given codebook $C$, with $\mathrm{d_H}(\bfc_{i_l},\bfc_{i_{l+1}}) = d_l, \, 1 \le l \le k_1-1$, $\mathrm{d_H}(\bfc_{i_{k_1}},\bfc_{i_{1}}) = d_{k_1}$, $\mathrm{d_H}(\bfc_{i_l},\bfc_{i_{l+1}}) = d_l, \, k_1+1 \le l \le k_1+k_2-1$, $\mathrm{d_H}(\bfc_{i_{k_1+k_2}},\bfc_{i_{k_1+1}}) = d_{k_1+k_2}$, we have $\mathrm{d_H}(C_{\pi_0}, C_{\sigma}) = \sum_{l=1}^{k_1+k_2} d_l$, and therefore
\begin{equation}
\Pr\{\pi_0 \to \sigma\} \le 2^{-\left(\sum_{l=1}^{k_1+k_2} d_l\right) \alpha_p} . \label{eq:ErrProb_two_kcycles}
\end{equation}
Combining \eqref{eq:WeakIndependence_v3} and \eqref{eq:ErrProb_two_kcycles}, we can upper bound $P_{\TRC,\sigma}$ by
\begin{align}
&\frac{1}{\alpha_n} \sum_{\substack{d_0 \le d_l \le n-d_0,\\1 \le l \le k_1+k_2}} 2^{-n \left( (\sum_{l=1}^{k_1+k_2} (d_l/n)\alpha_p)\right)} \nonumber\\
&~~~~~~~~~\times 2^{-n \left( \sum_{l=1}^{k_1-1} (1 - H(d_l/n)) + \sum_{l=k_1+1}^{k_1+k_2-1} (1 - H(d_l/n)) \right)} .
\end{align}
The above expression can be equivalently written as
\begin{equation}
\frac{1}{\alpha_n} \left(\eta_{k}\right)^2 \left(\zeta_{l}\right)^{k_1+k_2-2}, \label{eq:P_TRC_two_kcycles_v1}
\end{equation}
where $\zeta_{l}$ and $\eta_k$ are given by \eqref{eq:zeta_l_def} and \eqref{eq:eta_k_def}, respectively. Now, applying \eqref{eq:eta_k_def}, \eqref{eq:zeta_l_approx} in \eqref{eq:P_TRC_two_kcycles_v1} for $0 \le R < R_{\TRC}(p)$, we get
\begin{equation}
P_{\TRC,\sigma} \le \frac{1}{\alpha_n} 2^{-n \left( (k_1+k_2 - 2)(1 - H(\underline{\delta})) + (k_1+k_2) (\underline{\delta} \alpha_p - c_n) \right)},  \label{eq:P_TRC_two_kcycles_v2}
\end{equation}
where $\sigma = (i_1~i_2~\cdots~i_{k_1})(i_{k_1+1}~i_{k_1+2}~\cdots~i_{k_1+k_2})$. As $k_1 \ge 2$ and $k_2 \ge 2$, we have $2(k_1+k_2 - 2) \ge k_1+k_2$, and therefore for $~0 \le R < R_{\TRC}(p)$, we have
\begin{equation}
P_{\TRC,\sigma} \le \frac{1}{\alpha_n} 2^{-n(k_1+k_2)\left(0.5 (1 - H(\underline{\delta})) + \underline{\delta} \alpha_p - c_n \right)} . \label{eq:P_TRC_two_kcycles_v3}
\end{equation}

\subsubsection{General $\sigma \in S_m$ with $\sigma \neq \pi_0$}
If permutation $\sigma$ is a product of $r$ disjoint cycles of length $k_1, \ldots, k_r$, respectively, then similar to \eqref{eq:P_TRC_kcycle_v3}, \eqref{eq:P_TRC_two_kcycles_v3}, we have for $0 \le R \le R_{\TRC}(p)$, 
\begin{equation}
P_{\TRC,\sigma} \le \frac{1}{\alpha_n} 2^{-n\left(\sum_{t=1}^r k_t\right )\left(0.5 (1 - H(\underline{\delta})) + \underline{\delta} \alpha_p - c_n \right)}.\label{eq:P_TRC_general_sigma}
\end{equation}

\subsubsection{Putting it all together}
For $1 \le j \le m$, if we define $P_{\TRC,\Sigma_j} \triangleq \sum_{\sigma \in \Sigma_j} P_{\TRC,\sigma}$, where $\Sigma_j$ is given by \eqref{eq:Sigma_def}, then \eqref{eq:TRC_Opt_D_UnionBound} can be equivalently expressed as
\begin{equation}
\underline{D}(n,R,p) \le \sum_{j=2}^m P_{\TRC,\Sigma_j}. \label{eq:TRC_Opt_D_UnionBound_v2}
\end{equation}
If $\sigma$ is a product of $r$ disjoint cycles of length $k_1, \ldots, k_r$, respectively, and $s = \sum_{t=1}^r k_t$, then $\sigma$ belongs to the set $\Sigma_{s}$, and $P_{\TRC,\sigma}$ is given by \eqref{eq:P_TRC_general_sigma}. Equivalently, for a given $j \ge 2$, if $\sigma \in S_m$ belongs to the set $\Sigma_j$, then for $0 \le R < R_{\TRC}(p)$,
\begin{equation}
P_{\TRC,\sigma} \le \frac{1}{\alpha_n} 2^{-n j \left(0.5 (1 - H(\underline{\delta})) + \underline{\delta} \alpha_p - c_n \right)}. \label{eq:P_TRC_general_sigma_v2}
\end{equation}
The size of $\Sigma_j$ satisfies $|\Sigma_j| < \prod_{i=0}^{j-1} (m-i) < 2^{njR}$. Therefore, for $0 \le R < R_{\TRC}(p)$, we have
\begin{align}
P_{\TRC,\Sigma_j} &= \sum_{\sigma \in \Sigma_j} P_{\TRC,\sigma} \nonumber \\
&\le \frac{1}{\alpha_n} 2^{-n j \left(0.5 (1 - H(\underline{\delta})) + \underline{\delta} \alpha_p - c_n \right)} \, 2^{njR} \nonumber \\
&= \frac{1}{\alpha_n} 2^{-n j \left(0.5 (1 - H(\underline{\delta})) - R + \underline{\delta} \alpha_p - c_n \right)}. \label{eq:P_TRC_Sigma_j_v2}
\end{align}
Now, if we define $\beta_n \triangleq 2^{-n \left(0.5 (1 - H(\underline{\delta})) - R + \underline{\delta} \alpha_p - c_n \right)}$, then \eqref{eq:P_TRC_Sigma_j_v2} can be equivalently expressed as $P_{\TRC,\Sigma_j} \le (1/\alpha_n) \beta^j$. As $c_n = o(1)$, there exists $\hat{N}$ such that for $n \ge \hat{N}$, we have $c_n < 0.5 (1 - H(\underline{\delta})) - R + \underline{\delta} \alpha_p$ and hence $\beta_n < 1$. Therefore, for $n \ge \hat{N}$ and $0 \le R < R_{\TRC}(p)$, we have
\begin{align}
\underline{D}(n,R,p) &\le \frac{1}{\alpha_n} \sum_{j=2}^m \beta_n^j \nonumber \\
&< \frac{1}{\alpha_n} \frac{\beta_n^2}{\left(1 - \beta_n\right)} \nonumber \\
&\doteq \frac{\beta_n^2}{\left(1 - \beta_n\right)} \label{eq:TRC_doteq_i} \\
&\doteq \beta_n^2 \label{eq:TRC_doteq_ii} \\
&= 2^{-n \left(1 - H(\underline{\delta}) - 2R + 2 \underline{\delta} \alpha_p - 2 c_n \right)}  \nonumber \\
&\doteq 2^{-n \left(1 - H(\underline{\delta}) - 2R + 2 \underline{\delta} \alpha_p \right)}, \label{eq:TRC_doteq_iii}
\end{align}
where \eqref{eq:TRC_doteq_i} follows because $\alpha_n \to 1$ as $n \to \infty$~\cite{Barg02}, \eqref{eq:TRC_doteq_ii} follows because $\beta_n = o(1)$, and \eqref{eq:TRC_doteq_iii} follows because $c_n = o(1)$. Note that $\underline{\delta} = \delta_{\GV}(2R) - \epsilon$, and so $\lim_{\epsilon \to 0} \underline{\delta} = \delta_{\GV}(2R)$ and $\lim_{\epsilon \to 0} \left(1 - H(\underline{\delta}) - 2R + 2 \underline{\delta} \alpha_p \right) = 2 \delta_{\GV}(2R) \alpha_p$. As $\epsilon$ can be made arbitrarily small, it follows from \eqref{eq:TRC_doteq_iii} that for $0 \le R < R_{\TRC}(p)$, we have
\begin{equation}
\underline{D}(n,R,p) \dotleq 2^{-n \left(2 \delta_{\GV}(2R) \alpha_p \right)}.\label{eq:TRC_Distortion_UB_v2}
\end{equation}

The following theorem encapsulates the main result of this subsection on bounding the bee-identification exponent, $E_{\underline{D}}(R,p)$, using joint decoding for TRC.
\begin{theorem} \label{thm:TRC_Exp_JointDecoding}
	We have
	\begin{equation}
	E_{\underline{D}}(R,p) \ge 2\delta_{\GV}(2R)\,\alpha_p ,\ \ 0 \le R < R_{\TRC}(p). \label{eq:distortion_exponent_LB_v3}
	\end{equation}
\end{theorem}
\begin{IEEEproof}
	Follows from \eqref{eq:distortion_exponent} and \eqref{eq:TRC_Distortion_UB_v2}.
\end{IEEEproof}

We note that the above lower bound for $E_{\underline{D}}(R,p)$ using TRCs with joint barcode decoding is \emph{twice} the corresponding bound obtained using independent barcode decoding (see~\eqref{eq:TRC_ErrExp_LB}). The following proposition shows that the lower bound given by Thm.~\ref{thm:TRC_Exp_JointDecoding} using TRC is \emph{strictly better} than corresponding bound using RCE (see Thm.~\ref{thm:RCE_Exp_JointDecoding}) for $0 \le R < R_{\TRC}(p)$.

\begin{proposition} \label{prop:TRC_IsBetterThan_RCE}
The lower bound on $E_{\underline{D}}(R,p)$ in \eqref{eq:distortion_exponent_LB_v3} obtained for TRC is strictly better than the corresponding bound in \eqref{eq:identification-error_exponent_LB_JointDecoding} obtained for RCE when $0 \le R < R_{\TRC}(p)$.
\end{proposition}
\begin{IEEEproof}
It is known that $E_{\TRC}(R,p) > E_{\mathrm{r}}(R,p)$ when $0 \le R < R_{\TRC}(p)$ \cite{Barg02}. Further, using explicit numerical computation, it can be shown that $2R_0(p) \ge R_1(p) + 2 R_{\TRC}(p)$. Therefore, it follows that for $0 \le R < R_{\TRC}(p)$, we have
\begin{align*}
2\delta_{\GV}(2R)\,\alpha_p &= 2 \left(E_{\TRC}(R,p) - R\right) \\
&> 2 \left(E_{\mathrm{r}}(R,p) - R\right) = 2 (R_0(p) - 2R) \\
&\ge R_1(p) - 2R + 2(R_{\TRC}(p) - R) \\
&> R_1(p) - 2R \ge \eta_p(R).
\end{align*}
\end{IEEEproof}

The next section presents an explicit upper bound for $E_{\underline{D}}(R,p)$ which applies to all possible codebook designs. 
\section{Upper Bound on the Bee-Identification Exponent} \label{Sec:UB_IE_Exponent}
This section presents an upper bound on the bee-identification exponent $E_{\underline{D}}(R,p)$. Towards this, we define the following optimum minimum distance metrics
\begin{align*}
d^*(n,R) &\triangleq \max_{C \in \mathscr{C}(n,R)} \min_{\substack{\bfc_{i}, \bfc_{j} \in C\\ \bfc_i \neq \bfc_j}} \mathrm{d_H}(\bfc_i,\bfc_j) , \\
\delta^*(n,R) &\triangleq d^*(n,R)/n , \\
\delta^*(R) &\triangleq \limsup_{n \to \infty} \delta^*(n,R) .
\end{align*}
For any given codebook $C \in \mathscr{C}(n,R)$, we show that there exists a set $\mathscr{I}_{C}$ consisting of pairs of codeword indices $(i,j)$, $i \neq j $, with the following properties:
\begin{enumerate}[(i)]
\item If $(i,j) \in \mathscr{I}_{C}$, then $\mathrm{d_H}(\bfc_i,\bfc_j) \le d^*(n,R-\frac{1}{n})$.
\item If $(i,j) \in \mathscr{I}_{C}$ and $(\hat{\imath},\hat{\jmath}) \in \mathscr{I}_{C}$, then $\hat{\imath} \neq i, \hat{\imath} \neq j$ and $\hat{\jmath} \neq i, \hat{\jmath} \neq j$.
\item Size of set $\mathscr{I}_{C}$ is at least $m/4$. 
\end{enumerate}
A set satisfying the above properties can be constructed iteratively as follows.
\begin{itemize}
	\item \emph{Step 1}: For a given codebook $C \in \mathscr{C}(n,R)$, initialize $\mathscr{I}_{C}$ to be the empty set and let $\mathcal{T} = C$.
	\item \emph{Step 2}: As $\mathcal{T}$ contains at least $m/2$ codewords, there exists $\bfc_i, \bfc_j \in \mathcal{T}$, with $i \neq j$, satisfying $\mathrm{d_H}(\bfc_i,\bfc_j) \le d^*(n,R-\frac{1}{n})$. Include the pair $(i,j)$ to $\mathscr{I}_{C}$, and let $\mathcal{T} = \mathcal{T} \setminus \{\bfc_i,\bfc_j\}$.
	\item \emph{Step 3}: If $|\mathscr{I}_{C}| < m/4$, then go to \emph{Step 2}, else stop.
\end{itemize}

Let the receiver employ ML decoding, and interpret each pair $(i,j) \in \mathscr{I}_{C}$ as a transposition $\sigma = (i~j)$ that interchanges indices $i$ and $j$. Let $A_{(i,j)}$ denote the error event that the receiver incorrectly decodes the channel induced permutation to transposition $(i~j)$ (instead of the identity permutation $\pi_0$), i.e. $A_{(i,j)} = \{ \pi_0 \to (i~j)\}$.
Then, the bee-identification error probability $D(C,p,\phi)$ can be lower bounded as
\begin{equation}
D(C,p,\phi) \ge \Pr\left\{\displaystyle \bigcup_{(i,j) \in \mathscr{I}_{C}} A_{(i,j)} \right\}. \label{eq:Distortion_LB_v0}
\end{equation}
Using de Caen's lower bound on the probability of a union~\cite{DeCaen97}, the expression on the right side in \eqref{eq:Distortion_LB_v0} can itself be lower bounded by
\begin{align}
&~~\sum_{(i,j)\in \mathscr{I}_{C}} \frac{\left(\Pr\{A_{(i,j)}\}\right)^2}{\displaystyle \Pr\{A_{(i,j)}\} + \sum_{\substack{(\hat{\imath},\hat{\jmath}) \in \mathscr{I}_{C}\\ (\hat{\imath},\hat{\jmath}) \neq (i,j)}} \Pr\left\{A_{(i,j)}  \cap A_{(\hat{\imath},\hat{\jmath})}\right\}}, \nonumber \\
&\overset{(a)}{=} \sum_{(i,j)\in \mathscr{I}_{C}} \frac{\left(\Pr\{A_{(i,j)}\}\right)^2}{\displaystyle \Pr\{A_{(i,j)}\} + \sum_{\substack{(\hat{\imath},\hat{\jmath}) \in \mathscr{I}_{C}\\ (\hat{\imath},\hat{\jmath}) \neq (i,j)}} \Pr\left\{A_{(i,j)}\right\} \Pr\left\{A_{(\hat{\imath},\hat{\jmath})}\right\}}, \nonumber \\
&\ge~~ \frac{\displaystyle \sum_{(i,j)\in \mathscr{I}_{C}} \Pr\{A_{(i,j)}\}}{\displaystyle 1 + \sum_{(\hat{\imath},\hat{\jmath}) \in \mathscr{I}_{C}} \Pr\left\{A_{(\hat{\imath},\hat{\jmath})}\right\}} , \label{eq:Distortion_LB_v1}
\end{align}
where $(a)$ follows because events $A_{(i,j)}$ and $A_{(\hat{\imath},\hat{\jmath})}$ are independent when $(\hat{\imath},\hat{\jmath}) \neq (i,j)$. Now 
\begin{align}
\sum_{(i,j)\in \mathscr{I}_{C}} \Pr\{A_{(i,j)}\} &\overset{(b)}{\dotgeq} \sum_{(i,j)\in \mathscr{I}_{C}} 2^{-n\left(2\delta^*\left(n,R-\frac{1}{n}\right) \alpha_p\right)},  \nonumber \\
&\doteq \sum_{(i,j)\in \mathscr{I}_{C}} 2^{-n\left(2\delta^*(n,R) \alpha_p\right)}, \nonumber \\
&\overset{(c)}{\ge} 2^{-n\left(2\delta^*(n,R) \alpha_p - \left(R - \frac{2}{n}\right) \right)}, \nonumber \\
&\doteq 2^{-n\left(2\delta^*(R) \alpha_p - R \right)} , \label{eq:Prob_Aij_Sum_LB}
\end{align}
where $(b)$ follows from the fact that $\mathrm{d_H}(C_{\pi_0}, C_{(i,j)}) \le  2\, d^*(n,R-\frac{1}{n})$ for $(i,j) \in \mathscr{I}_{C}$, and $(c)$ follows because $|\mathscr{I}_{C}| \ge m/4$.  If $R_{\UB}(p) \triangleq \sup \{R : 2\delta^*(R) \alpha_p > R \}$, then combining \eqref{eq:Distortion_LB_v0}, \eqref{eq:Distortion_LB_v1}, \eqref{eq:Prob_Aij_Sum_LB}, and noting that $x/(1+x)$ increases with $x$, we have
\begin{align}
D(C,p,\phi) &\dotgeq \frac{2^{-n\left(2\delta^*(R) \alpha_p - R\right)}}{1+2^{-n\left(2\delta^*(R) \alpha_p - R\right)}}, \nonumber \\
&\doteq 2^{-n\left(2\delta^*(R) \alpha_p - R\right)}, \ \ 0 \le R < R_{\UB}(p) . \label{eq:Identification-error_LB_v3}
\end{align}
As \eqref{eq:Identification-error_LB_v3} is true for all $C \in \mathscr{C}(n,R)$, we have 
\begin{equation}
\underline{D}(n,R,p) \dotgeq 2^{-n\left(2\delta^*(R) \alpha_p - R\right)}, \ \ 0 \le R < R_{\UB}(p) . \label{eq:OptIdentification-error_LB}
\end{equation}
The value $\delta^*(R)$ can be upper bounded as~\cite{McEliece77,Litsyn99}
\begin{equation}
\delta^*(R) \le \delta_{\LP}(R) \triangleq \frac{1}{2} - \sqrt{ \delta_{\GV}(1-R) (1-\delta_{\GV}(1-R)) }. \label{eq:delta_star_UB}
\end{equation}

The following theorem provide an upper bound on the bee-identification exponent $E_{\underline{D}}(R,p)$.
\begin{theorem}
	We have
	\begin{equation}
	E_{\underline{D}}(R,p) \le |2\delta^*(R) \alpha_p - R|^+ \le |2\delta_{\LP}(R) \alpha_p - R|^+ . \label{eq:Identification-errorExponent_UB}
	\end{equation}
\end{theorem}
\begin{IEEEproof}
	Follows immediately from \eqref{eq:OptIdentification-error_LB} and \eqref{eq:delta_star_UB}.
\end{IEEEproof}

The following corollary shows that $E_{\underline{D}}(R,p)$ can be explicitly characterized with a rather simple expression when rate $R$ tends to zero.
\begin{corollary} \label{cor:ExactBeeExponentWhenReq0}
	We have
	\begin{equation}
	\lim_{R \to 0} E_{\underline{D}}(R,p) = \alpha_p. \label{eq:ExactBeeExponentWhenReq0}
	\end{equation}
\end{corollary}
\begin{IEEEproof}
As $\lim_{R \to 0} \delta_{\LP}(R) = 0.5$, we have from \eqref{eq:Identification-errorExponent_UB} that
\begin{equation}
\lim_{R \to 0} E_{\underline{D}}(R,p) \le \lim_{R \to 0} \left(2\delta_{\LP}(R) \alpha_p - R\right) = \alpha_p. \label{eq:ErrExpUB_Req0}
\end{equation}
On the other hand, we have $\lim_{R \to 0} \delta_{\GV}(R) = 0.5$ and so it follows from \eqref{eq:distortion_exponent_LB_v3} that
\begin{equation}
\lim_{R \to 0} E_{\underline{D}}(R,p) \ge \lim_{R \to 0} 2\delta_{\GV}(2R) \alpha_p = \alpha_p. \label{eq:ErrExpLB_Req0}
\end{equation}
The proof is completed by using \eqref{eq:ErrExpUB_Req0} and \eqref{eq:ErrExpLB_Req0}.
\end{IEEEproof}
The above corollary shows that the lower bound on $E_{\underline{D}}(R,p)$ given by \eqref{eq:distortion_exponent_LB_v3}, and the upper bound on $E_{\underline{D}}(R,p)$ given by \eqref{eq:Identification-errorExponent_UB} become \emph{tight} as $R \to 0$.

\section{A Numerical Example}
\begin{figure}[t]
	\centering
	\captionsetup{justification=raggedright,singlelinecheck=false,width=0.49\textwidth}
	\includegraphics[width=0.49\textwidth]{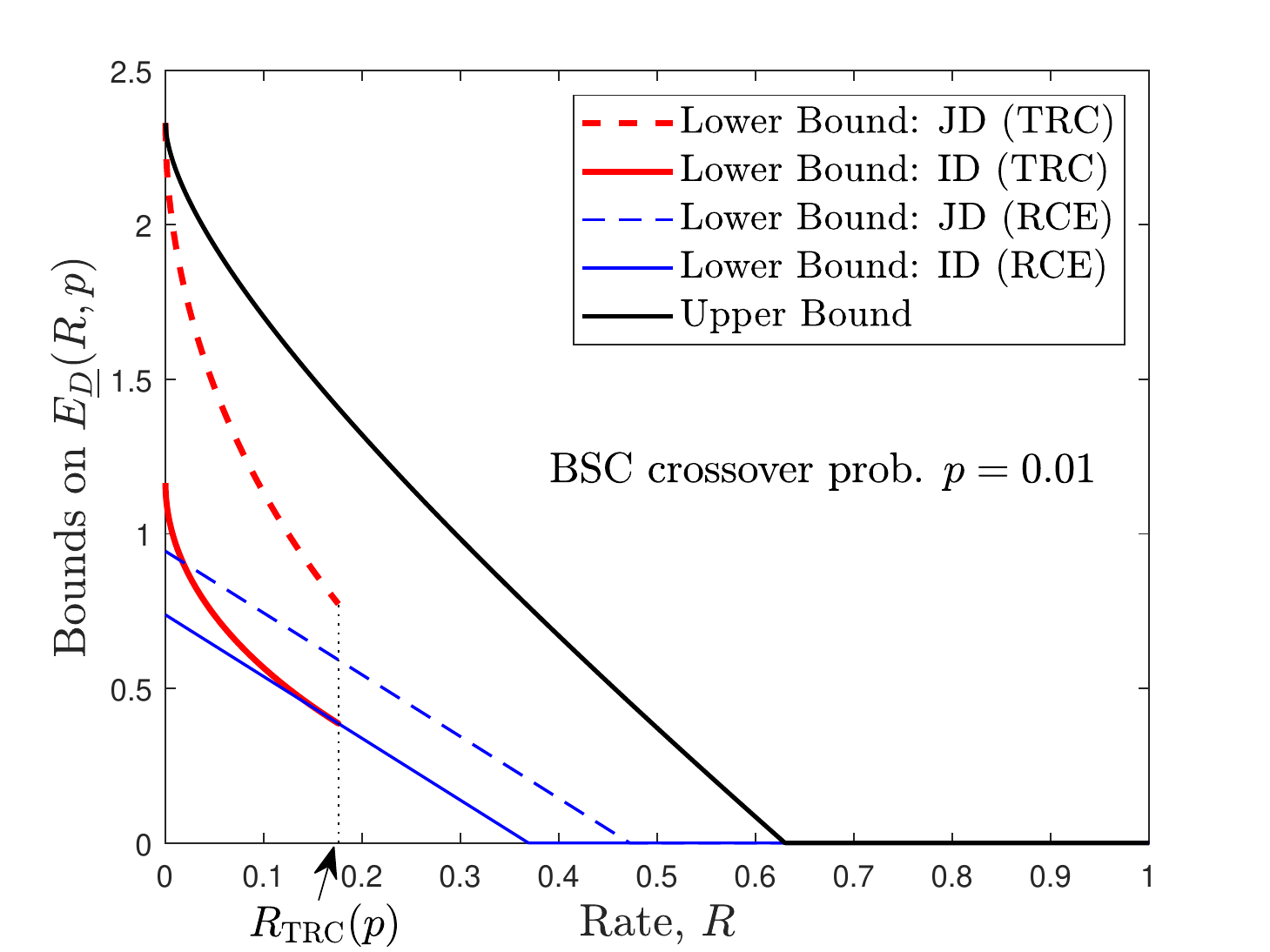}
	\caption{Lower bounds on $E_{\underline{D}}(R,p)$ with independent decoding (ID) and joint decoding (JD) using TRC and RCE.~The upper bound holds for all codebook designs.}
	\label{Fig:ErrExp_vs_R}
\end{figure}

Fig.~\ref{Fig:ErrExp_vs_R} plots different bounds for the bee-identification exponent $E_{\underline{D}}(R,p)$. The explicit lower bound for RCE with independent decoding (ID) (respectively, joint decoding (JD)) is given by \eqref{eq:identification-error_exponent_LB_v2} (respectively, \eqref{eq:identification-error_exponent_LB_JointDecoding}). The performance with JD is seen to be much better than with ID. When $0 \le R < R_{\TRC}(p)$, the explicit lower bound for TRC with ID (respectively, JD) is given by \eqref{eq:TRC_ErrExp_LB} (respectively, \eqref{eq:distortion_exponent_LB_v3}). As shown in Prop.~\ref{prop:TRC_IsBetterThan_RCE}, the lower bound obtained using TRC with joint decoding is better than the corresponding bound using RCE. The upper bound is given by \eqref{eq:Identification-errorExponent_UB} and holds for all possible codebook designs. Further, as shown in Cor.~\ref{cor:ExactBeeExponentWhenReq0}, it is observed from Fig.~\ref{Fig:ErrExp_vs_R} that $\lim_{R \to 0} E_{\underline{D}}(R,p) = \alpha_p = 2.33$ for $p=0.01$.

\section{Discussion}
We introduced the information-theoretic ``bee-identification problem'' which arises naturally in different massive identification settings. We derived explicit upper and lower bounds on the bee-identification exponent, and showed that joint decoding of barcodes provides a significantly better exponent than separate decoding followed by permutation inference. For low rates, we showed that the lower bound on the bee-identification exponent obtained using TRC is strictly better than the corresponding bound obtained using RCE.  Moreover, when the rate approaches zero, we showed that the upper bound on the bee-identification exponent coincides with the lower bound obtained using TRC with joint barcode decoding.

Relative to the independent decoding of barcodes, the performance improvement with joint decoding comes at a cost of increased computational complexity. For joint decoding, an exhaustive search entails comparing the received noisy \& permuted version of the codebook with $m!$ row-permutations of the codebook. This may be computationally prohibitive even for moderate values of blocklength $n$ when $m$ scales exponentially with $n$. In practice, intermediate performance between the extremes of independent decoding and joint decoding may be achieved with manageable complexity using ideas from generalized minimum distance decoding~\cite{Forney66_GMD}. In particular, the decoding process may proceed in two steps: The first step involves independent decoding of each barcode where an erasure is declared if the distance between the received noisy barcode to the nearest barcode in the codebook exceeds a threshold. The second step fixes the codebook row-indices corresponding to the un-erased barcodes, and then decodes the erased barcodes by jointly comparing their received noisy version to different row-permutations of the codebook corresponding to the non-fixed indices. This results in significant reduction in complexity in case only a few barcodes are declared as erasure in the first step. Therefore, we have a tradeoff between performance and complexity via an appropriate choice of the distance threshold parameter for declaring an erasure.

The work in this paper may be extended by considering different variants of the bee-identification error metric, for instance, where error is flagged only when the fraction of incorrectly decoded barcodes exceeds a threshold. Another interesting scenario for future analysis is the problem formulation where some of the $m$ rows in codebook $C$ are deleted, due to some bees being outside the hive when taking the picture.

\appendices

\section{Proof of Prop.~\ref{prop:DistanceIndependence_RCE}} \label{app:DistanceIndependence_RCE}
\begin{IEEEproof}
	Let $\gamma_{k-1}, \tilde{\gamma}_{k-1} \in \mathbb{F}_{2^n}$, and $\Delta \triangleq \gamma_{k-1} \oplus \tilde{\gamma}_{k-1}$, where $\oplus$ denotes modulo-2 addition. Then, $\Pr\{\mathrm{d_H}(\gamma_{k-1},\bfc_{k}) = d_{k-1}\} =  \Pr\{\mathrm{d_H}(\tilde{\gamma}_{k-1},\bfc_{k}+\Delta) = d_{k-1}\} \overset{(\mathrm{i})}{=} \Pr\{\mathrm{d_H}(\tilde{\gamma}_{k-1},\bfc_{k}) = d_{k-1}\}$, where $(\mathrm{i})$ follows from the fact that for a given $\Delta$, the distribution of $\bfc_{k}+\Delta$ is same as the distribution of $\bfc_{k}$. This implies that	
$\Pr\{\mathrm{d_H}(\bfc_{k-1},\bfc_{k}) = d_{k-1} | \bfc_{k-1} = \gamma_{k-1}\} \overset{(\mathrm{ii})}{=} \Pr\{\mathrm{d_H}(\bfc_{k-1},\bfc_{k}) = d_{k-1}\}$. Then $\Pr\{ \bigcap_{i=1}^{k-1} \{\mathrm{d_H}(\bfc_i,\bfc_{i+1}) = d_i \}\}$ can be expressed as
\begin{align}
&\sum_{\gamma_1,\ldots,\gamma_{k-1} \in \mathbb{F}_{2^n}} \Bigg( \Pr\left\{ \bigcap_{i=1}^{k-1} \{\bfc_i = \gamma_i\}\right\} \nonumber \\
&~~~~~~~\times \Pr\left\{ \bigcap_{i=1}^{k-1} \left\{\mathrm{d_H}(\bfc_i,\bfc_{i+1}) = d_i \right\} | \bigcap_{i=1}^{k-1} \{\bfc_i = \gamma_i\} \right\} \Bigg), \nonumber \\
&= \sum_{\gamma_1,\ldots,\gamma_{k-1}} \Bigg( \Pr\left\{ \bigcap_{i=1}^{k-1} \{\bfc_i = \gamma_i\}\right\} \one_{\left\{ \bigcap_{i=1}^{k-2} \left\{\mathrm{d_H}(\gamma_i,\gamma_{i+1}) = d_i \right\} \right\}} \nonumber \\
&~~~~~~~~~~~~~~~~\times \Pr\left\{ \mathrm{d_H}(\bfc_{k-1},\bfc_{k}) = d_{k-1} | \bfc_{k-1} = \gamma_{k-1} \right\} \Bigg), \nonumber \\
&\overset{(\mathrm{iii})}{=} \sum_{\gamma_1,\ldots,\gamma_{k-1}} \Bigg(\Pr\left\{ \bigcap_{i=1}^{k-1} \{\bfc_i = \gamma_i\}\right\} \one_{\left\{ \bigcap_{i=1}^{k-2} \left\{\mathrm{d_H}(\gamma_i,\gamma_{i+1}) = d_i \right\}\right\}} \nonumber \\
&~~~~~~~~~~~~~~~~~~~~~~~\times \Pr\left\{ \mathrm{d_H}(\bfc_{k-1},\bfc_{k}) = d_{k-1} \right\} \Bigg), \nonumber \\
&=\Pr\Big\{\bigcap_{i=1}^{k-2} \mathrm{d_H}(\bfc_i,\bfc_{i+1})= d_i \Big\} \Pr\left\{ \mathrm{d_H}(\bfc_{k-1},\bfc_{k}) = d_{k-1} \right\},\label{eq:independence_of_distances_v3}
\end{align}
where $\one_{\{\cdot\}}$ denotes the indicator function, and $(\mathrm{iii})$ follows from $(\mathrm{ii})$. Recursively applying \eqref{eq:independence_of_distances_v3}, we get
\begin{equation*}
\Pr\left\{ \bigcap_{i=1}^{k-1} \left\{\mathrm{d_H}(\bfc_i,\bfc_{i+1}) = d_i \right\}\right\} = \prod_{i=1}^{k-1} \Pr\left\{\mathrm{d_H}(\bfc_i,\bfc_{i+1}) = d_i \right\}.
\end{equation*}
Now, \eqref{eq:independence_of_distances_v1} follows from the fact that $\Pr\left\{\mathrm{d_H}(\bfc_i,\bfc_{i+1}) = d_i \right\} \le 2^{-n(1-H(d_i/n))}$ when $\bfc_i$ and $\bfc_{i+1}$ are uniformly distributed over $\mathbb{F}_{2^n}$~\cite{Barg02}.
\end{IEEEproof}

\section{Proof of Prop.~\ref{prop:WeakIndependence_TRC}} \label{app:WeakIndependence_TRC}
\begin{IEEEproof}
For $1 \le i \le m=2^{nR}$, let $\bfc_{i}$ denote the $i$-th row of codebook $C$. Let $\mathbb{F}_{2^n}$ denote the space of all $n$-length binary vectors, and let $\gamma_i \in \mathbb{F}_{2^n}$ for $1 \le i \le m$. Let $Q_{\TRC}\left\{ \bigcap_{i=1}^{m} \{\bfc_i = \gamma_i\}\right\}$ denote the probability $\Pr\left\{ \bigcap_{i=1}^{m} \{\bfc_i = \gamma_i\}\right\}$ when $C$ is uniformly distributed over $\sC_{\TRC}(n,R)$. Then, we have
\begin{align}
&Q_{\TRC}\left\{ \bigcap_{i=1}^{m} \{\bfc_i = \gamma_i\}\right\} \nonumber \\
&= \frac{1}{\alpha_n} Q_{\RCE}\left\{ \bigcap_{i=1}^{m} \{\bfc_i = \gamma_i\}\right\} \one_{\left\{(\gamma_1, \gamma_2, \ldots, \gamma_m) \in \sC_{\TRC}(n,R)\right\}}, \label{eq:TRC_to_RCE}
\end{align}
where $\one_{\{\cdot\}}$ denotes the indicator function. Further, let $Q_{\RCE}\left\{ \bigcap_{l=1}^{k-1} \left\{\mathrm{d_H}(\bfc_{i_l},\bfc_{i_{l+1}}) = d_l\right\} \right\}$ denote the probability $\Pr\left\{ \bigcap_{l=1}^{k-1} \left\{\mathrm{d_H}(\bfc_{i_l},\bfc_{i_{l+1}}) = d_l\right\} \right\}$ when codebook $C$ is uniformly distributed over $\sC(n,R)$. Then,
\begin{align*}
&Q_{\TRC}\left\{ \bigcap_{l=1}^{k-1} \left\{\mathrm{d_H}(\bfc_{i_l},\bfc_{i_{l+1}}) = d_l\right\} \right\} \\
&= \sum_{\substack{\gamma_i \in \mathbb{F}_{2^n},\\1 \le i \le m}} Q_{\TRC}\left\{ \bigcap_{i=1}^{m} \{\bfc_i = \gamma_i\}\right\} \one_{\left\{ \bigcap_{l=1}^{k-1} \mathrm{d_H}(\gamma_{i_l},\gamma_{i_{l+1}}) = d_l\right\}},\\
&\overset{(a)}{\le} \frac{1}{\alpha_n} \sum_{\substack{\gamma_i \in \mathbb{F}_{2^n},\\1 \le i \le m}} Q_{\RCE}\left\{ \bigcap_{i=1}^{m} \{\bfc_i = \gamma_i\}\right\} \one_{\left\{ \bigcap_{l=1}^{k-1} \mathrm{d_H}(\gamma_{i_l},\gamma_{i_{l+1}}) = d_l\right\}},\\
&= \frac{1}{\alpha_n} Q_{\RCE}\left\{ \bigcap_{l=1}^{k-1} \left\{\mathrm{d_H}(\bfc_{i_l},\bfc_{i_{l+1}}) = d_l\right\} \right\},\\
&\overset{(b)}{\le} \frac{1}{\alpha_n} \prod_{l=1}^{k-1} 2^{-n(1 - H(d_l/n))},
\end{align*}
where $(a)$ follows from \eqref{eq:TRC_to_RCE}, and $(b)$ follows from Prop.~\ref{prop:DistanceIndependence_RCE}.

\end{IEEEproof}

\balance 

\section*{Acknowledgement}
The authors acknowledge discussions with Ting-Yi Wu and Tim Gernat on the bee-identification problem formulation.

%\bibliographystyle{IEEEtran}
%\bibliography{abrv,conf_abrv,mybibfile}

% Generated by IEEEtran.bst, version: 1.13 (2008/09/30)

\end{document}